\documentclass[fleqn,twoside,twocolumn,nofootinbib,showkeys,a4paper]{revtex4} 
\usepackage[sec,nocpr]{ujp} 

\begin{document}
\title[Investigation of the Bosonic Spectrum]
{INVESTIGATION OF THE BOSONIC\\ SPECTRUM OF TWO-DIMENSIONAL
OPTICAL\\
GRAPHENE-TYPE LATTICES. SUPERFLUID PHASE}%
\author{I.V.~Stasyuk}
\affiliation{Institute for Condensed Matter Physics, Nat. Acad. of
Sci. of
Ukraine}
\address{1, Svientsitskii Str., Lviv 79011, Ukraine}
\author{O.V.~Velychko}%
\affiliation{Institute for Condensed Matter Physics, Nat. Acad. of
Sci. of
Ukraine}%
\address{1, Svientsitskii Str., Lviv 79011, Ukraine}%
\author{I.R.~Dulepa}
\affiliation{Institute for Condensed Matter Physics, Nat. Acad. of
Sci. of
Ukraine}
\address{1, Svientsitskii Str., Lviv 79011, Ukraine}

\udk{538.911, 538.941}%
\razd{\secvii}

\autorcol{I.V.\hspace*{0.7mm}Stasyuk, O.V.\hspace*{0.7mm}Velychko,
I.R.\hspace*{0.7mm}Dulepa}

\setcounter{page}{993}%

\begin{abstract}
The energy spectrum of a system of Bose atoms in the superfluid
phase in an optical lattice of the graphene type has been studied.
The dispersion laws for the energy bands and the single particle
spectral densities are calculated in the random phase approximation
and in the framework of the hard-core boson formalism, and their
changes at the transition from the normal phase to the superfluid
one are described.\,\,As a result of this transformation, the number
of subbands doubles.\,\,In the case of the subband energetic
equivalence, the Dirac points in the spectrum survive, and their
number becomes twice as much.\,\,When the subbands are energetically
nonequivalent, the Dirac points are absent.\,\,The shape of spectral
densities is shown to be sensitive to the changes in the temperature
and the chemical potential position.
\end{abstract}


\maketitle

\section{Introduction}

This work continues our researches dealing with the calculation of
the energy spectrum and one-particle spectral densities for a system
of Bose atoms in a two-dimensional honeycomb optical lattice of the
graphene type.\,\,Our previous results were reported in work
\cite{Stasyuk}.\,\,Unlike the problem of the electron spectrum in
graphene, we consider particles that are described by the non-Fermi
statistics.\,\,During the last time, a considerable attention was
attracted to the study of the thermodynamics and the energy spectrum
features of Bose particles in optical lattices.\,\,Interesting and
important is the problem concerning the spectrum modifications at
the phase transition associated with the Bose--Einstein condensation
of particles (the transition from the normal, NO, phase to the
superfluid, SF, one).

The Bose--Einstein (BE) condensation of bosonic atoms (Rb$^{87}$) in
the optical lattices formed as a result of the interference between
counter-propagating laser beams was observed for the first time in
2002 \cite{Greiner02a,Greiner02b}.\,\,Since that time, this effect
and the accompanying phenomena have been studied very
intensively.\,\,For optical lattices of the graphene type, the
transition to the SF phase was obtained experimentally in work
\cite{arr4}, where the regions of existence for various phases
depending on the values of chemical potential and parameters that
characterize short-range interactions and the particle dynamics were
found.\,\,The theoretical consideration in works
\cite{ar2,ar3,ar4,ar5} concerned the construction and the analysis
of phase diagrams, proceeding from the Bose--Hubbard model
\cite{Fisher89,Jaksch98}, which is generally adopted for the
description of a system of Bose atoms in optical lattices.\,\,The
energy spectrum of bosons in the graphene-type lattices was analyzed
in works \cite{a6,a7}.\,\,Issues concerning the spectrum topology
and the arrangement of Dirac points in the NO phase were considered,
but the spectrum modifications at the transition to the SF phase
were not analyzed.

An important influence on the formation of a spectrum of the system
of Bose atoms in the graphene-type lattice is exerted by short-range
correlations between particles, (in particular, the on-site
repulsion interaction $U$).\,\,Another complication in comparison
with the case of the graphene lattice is the energetic
nonequivalence of sublattices (it can be easily controlled by
changing the phases of laser beams that generate the optical
lattice).\,\,As was shown in work \cite{Stasyuk}, all those factors
give rise to substantial differences between the single particle
spectrum in the NO phase and the standard spectrum of
graphene.\,\,In particular, it was found that Dirac points in the
spectrum do not exist at arbitrary particle densities and,
accordingly, chemical potential values.\,\,In particular, they
disappear, if the chemical potential in the NO phase is located in
the gap between the energy subbands (such a gap emerges owing to the
energetic nonequivalence of sublattices).

In this work, in addition to calculations carried out in work
\cite{Stasyuk}, we consider the case where the system is in the
phase with the BE condensate (the SF phase).\,\,Our description is
based, as was done in work \cite{Stasyuk}, on the two-sublattice
model of hard-core bosons \cite{Whitlock63}, which is a limiting
case ($U\rightarrow\infty$) of the Bose--Hubbard model and is valid
for the low-population levels ($0\leq n\leq1$) of on-site states.
The general scheme to find Green's single particle functions for
this model in its pseudo-spin formulation is well-known.\,\,In this
work, we use the approach described in work \cite{a1}.\,\,We aim at
studying the special features in the reconstruction of the bosonic
band spectrum and spectral densities at the transition from the NO
phase to the SF one that occurs at the variations of the chemical
potential, energetic difference between the sublattices, and
temperature.

\section{Excitation\\ Spectrum in the Superfluid Phase}

If the intersite  interaction between particles is neglected, the
Hamiltonian of the lattice gas consisted of hard-core bosons looks
like
\begin{equation}
H=-\sum_{\langle ij\rangle}tb_{i}^{+}b_{j}+\sum_{i}(\varepsilon_{0}-\mu)n_{i},
\label{Pr-ista2.1}%
\end{equation}
where $t$ is the transfer integral between the nearest sites,
$\varepsilon _{0}$ the on-site energy of the particle, and $\mu$ the
chemical potential.\,\,Since $b_{i}^{+}$ and $b_{i}$ are the Pauli
operators, Hamiltonian (\ref{Pr-ista2.1}) can be rewritten in the
pseudo-spin representation with the use of the
transformations\vspace*{-2mm}
\begin{equation}
b_{i}^{+}=S_{i}^{-},\quad b_{i}=S_{i}^{+},\quad b_{i}^{+}b_{i}=n_{i}%
=\frac{1}{2}-S_{i}^{z}. \label{Pr-ista2.2}%
\end{equation}
In the two-sublattice case, $i\rightarrow(n,\alpha)$, where
$\alpha=A,B$ is the sublattice index.\,\,As a result,
\begin{equation}
\begin{array}{l}
\displaystyle\hat{H}=
 -\sum_{\langle n,n' \rangle}
            \biggl[
            J_{nn'}^{AB}(S_{nA}^x S_{n'B}^x\,
            +\\[3mm]
\displaystyle+\,J_{nn'}^{BA}(S_{nB}^x S_{n'A}^x
            +S_{nB}^y S_{n'A}^y)
            \biggr]-\\[3mm]
       \displaystyle -\,h_A \sum_n S_{nA}^z - h_B \sum_{n} S_{nB}^z\\[5mm]
        \displaystyle(J_{\langle n,n'\rangle}^{AB}=J_{\langle n',n\rangle}^{BA}=t,
        \quad
        h_{\alpha}=\varepsilon_{\alpha}-\mu).
        \label{Pr-ista2.3}
\end{array}
\end{equation}

In the phase with a Bose condensate, the order parameter is the
non-zero average
\begin{equation}
\begin{array}{l}
      \displaystyle S_{n\alpha}^z
      =
      \sigma_{n\alpha}^z \cos\vartheta_\alpha
      +
      \sigma_{n\alpha}^x \sin\vartheta_\alpha,\\[3mm]
      S_{n\alpha}^x
      =
      \displaystyle\sigma_{n\alpha}^x \cos\vartheta_\alpha
      -
      \sigma_{n\alpha}^z \sin\vartheta_\alpha,\\[3mm]
      S_{n\alpha}^y
      =
      \sigma_{n\alpha}^y,
      \label{Pr-ista2.4}
\end{array}
\end{equation}
where the angles $\vartheta_{\alpha}$ are determined by diagonalizing the
mean-field Hamiltonian
\begin{equation}
H_{{\mathrm{MF}}}=-\sum_{n\alpha}E_{\alpha}\sigma_{n\alpha}^{z}.
\label{Pr-ista2.5}%
\end{equation}
Since $\langle\sigma_{n\alpha}^{z}\rangle\neq0$ and $\langle\sigma_{n\alpha
}^{x}\rangle=\langle\sigma_{n\alpha}^{y}\rangle=0$, we obtain $\langle
S_{\alpha}^{x}\rangle=-\langle\sigma_{\alpha}^{z}\rangle\sin\vartheta_{\alpha
}$.

In the normal phase,\vspace*{-2mm}
\begin{equation}
\sin\vartheta_{\alpha}=0,\quad E_{\alpha}=h_{\alpha},\quad\langle
\sigma_{n\alpha}^{z}\rangle=\frac{1}{2}\tanh\frac{\beta
h_{\alpha}}{2}.
\label{Pr-ista2.6}%
\end{equation}
For the SF phase, the internal fields $E_{\alpha}$ and angles
$\vartheta_{\alpha}$ can be determined from the system of equations \cite{a1}%
\begin{equation}
\begin{array}{l}
      \displaystyle
    \sin^{2}\vartheta_{\alpha}
    =
    \frac{\langle \sigma_{\alpha}^{z}\rangle^{2}\langle \sigma_{\beta}^{z}\rangle^{2} J^{4}(0)-h_{\alpha}^{2}h_{\beta}^{2}}
    {\langle \sigma_{\alpha}^{z}\rangle^{2}J^{2}(0)[h_{\alpha}^{2}+\langle
        \sigma_{\beta}^{z}\rangle^{2}J^{2}(0)]},\\[5mm]
   \displaystyle \langle \sigma_{\alpha}^{z}\rangle
    =
    \frac12\tanh \frac{\beta E_{\alpha}}{2},\\[3mm]
 \displaystyle E_{\alpha}
    =
    \langle\sigma_{\alpha}^{z}\rangle J(0)\frac{\sqrt{h_{\alpha}^{2}+\langle\sigma_{\beta}^{z}\rangle^{2} J^{2}(0)}}
    {\sqrt{h_{\beta}^{2}+\langle\sigma_{\alpha}^{z}\rangle^{2}J^{2}(0)}},
\end{array}\label{Pr-ista2.7}
\end{equation}
and the transition from one phase into another one occurs when the condition
\begin{equation}
h_{A}h_{B}=\langle\sigma_{A}^{z}\rangle\langle\sigma_{B}^{z}\rangle
J^{2}(0)\equiv\langle\sigma_{A}^{z}\rangle\langle\sigma_{B}^{z}\rangle9t^{2}
\label{Pr-ista2.8}%
\end{equation}
is satisfied.\,\,Relation (\ref{Pr-ista2.8}) determines the
boundaries of regions for the NO and SF phases on the phase
plane.\,\,In Fig.\,\,1, the corresponding phase diagrams on the
plane $(T,h)$ \cite{Stasyuk,a1}, where $h=\frac {h_{A}+h_{B}}{2}$,
are exhibited for various values of parameter
$\delta=\frac{h_{A}-h_{B}}{2}$.

Green's two-time function
\begin{equation}
\langle\langle b_{l\alpha}|b_{n\beta}^{+}\rangle\rangle=\langle\langle
S_{l\alpha}^{+}|S_{n\beta}^{-}\rangle\rangle\equiv G_{l\alpha,n\beta}^{+-},
\label{Pr-ista2.9}%
\end{equation}
which can be used to obtain the bosonic spectrum, was found in works
\cite{Stasyuk,a1}, by using the method of equations of motion in the
random phase approximation.\,\,In the momentum-frequency
representation, its Fourier transform looks like
\[
   \langle\langle  S_{\mathrm{A}}^{+}|S_{\mathrm{A}}^{-}\rangle\rangle_{q,w}
    =
    \frac{\hbar}{2\pi}\langle\sigma_{\mathrm{A}}^{z}\rangle
    P_{q}^{\mathrm{A}}
    \biggl[
        (\hbar^{2}\omega^{2}{-}E_{\mathrm{A}}^{2})
        (\hbar^{2}\omega^{2}{-}E_{\mathrm{B}}^{2})\,-
    \]\vspace*{-7mm}
\begin{equation}
        -\,2M_{q}\hbar^{2}\omega^{2}
        -2N_{q}E_{\mathrm{A}}E_{\mathrm{B}}
        +M_{q}^{2}
    \biggr]^{-1}\!,
    \label{Pr-ista2.10}
\end{equation}
where the numerator of Green's function equals
\[
    P_{q}^{\mathrm{A}}(\hbar\omega)
    =
    \left[
        E_{\mathrm{A}}
        \left(\cos^{2}\vartheta_{\mathrm{A}}+1\right)
        +2\hbar\omega\cos\vartheta_{\mathrm{A}}\right]\times
\]\vspace*{-7mm}
  \begin{equation}
    \times
    \left(\hbar^{2}\omega^{2}-E_{\mathrm{B}}^{2}\right)
    -
    2\hbar\omega M_{q}\cos\vartheta_{\mathrm{A}}
    +
    \tilde{\Phi}_{q}^{\mathrm{A}}E_{\mathrm{B}}
\label{Pr-ista2.11}
\end{equation}
and the following notations are introduced:
\begin{equation}
\begin{array}{l}
      \displaystyle
    M_{q}
    =
    \Phi_{q}\cos\vartheta_{\mathrm{A}}\cos\vartheta_{\mathrm{B}},
    \\[1mm]
     \displaystyle N_{q}
    =
    \frac12\Phi_{q}
    \left(
        1+\cos^{2}\vartheta_{\mathrm{A}}\cos^{2}\vartheta_{\mathrm{B}}
    \right)\!,
    \\[3mm]
     \displaystyle \tilde{\Phi}_{q}^{\mathrm{A}}
    =
    \Phi_{q}\cos^{2}\vartheta_{\mathrm{A}}
    \left(1+\cos^{2}\vartheta_{\mathrm{B}}\right)\!,
     \\[2mm]
    \Phi_{q}
   =
    \displaystyle \langle\sigma_{A}^{z}\rangle\langle\sigma_{B}^{z}\rangle
        J^{2}(q).
    \label{Pr-ista2.12}
\end{array}
\end{equation}
Here,\vspace*{-3mm}
\begin{equation}
\begin{array}{l}
      \displaystyle
    J(\mathbf{q})=t\left[
        {\rm e}^{\mathrm{i}q_{y}a}+2{\rm e}^{-\mathrm{i}q_{y}a}\cos\biggl(\!q_{x}\frac{\sqrt{3}a}{2}\!\biggr)
    \!\right]\!,
   \\[4mm]
    J(0)=3t,
    \label{Pr-ista2.13}
\end{array}
\end{equation}
and $a$ is the distance between the nearest neighbors in the honeycomb lattice.

\begin{figure}
\vskip1mm
   \includegraphics[width=\columnwidth]{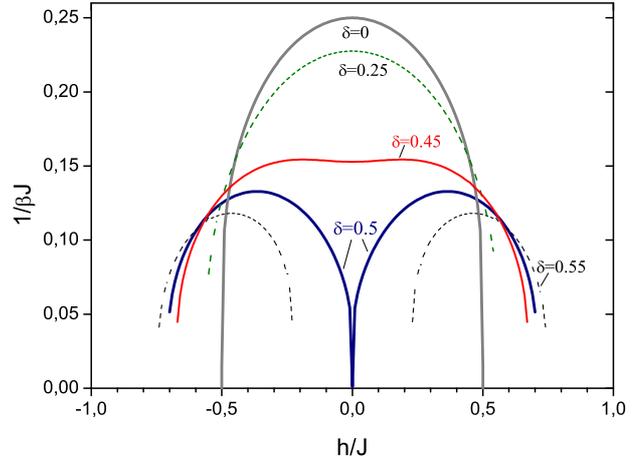}
\vskip-3mm    \caption{Phase diagrams in the $(T,h)$ plane at
$\delta=0$, 0.25, 0.45, 0.5, and 0.55 {\cite{a1}}}
    \label{fig01}
\end{figure}

\begin{figure*}[!t]
\strut%
\hfill%
\includegraphics[width=5.5cm]{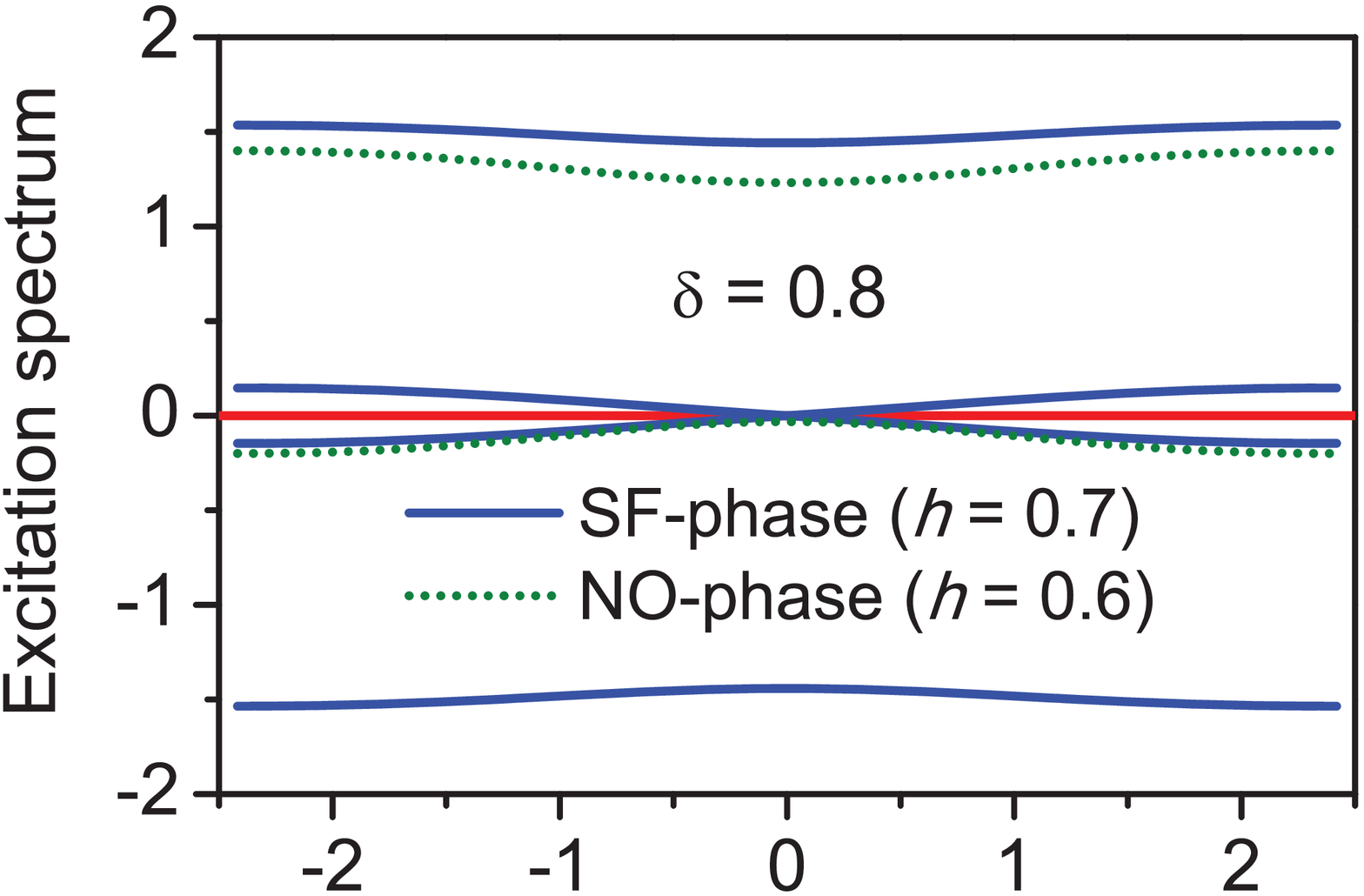}%
\hfill%
\includegraphics[width=5.5cm]{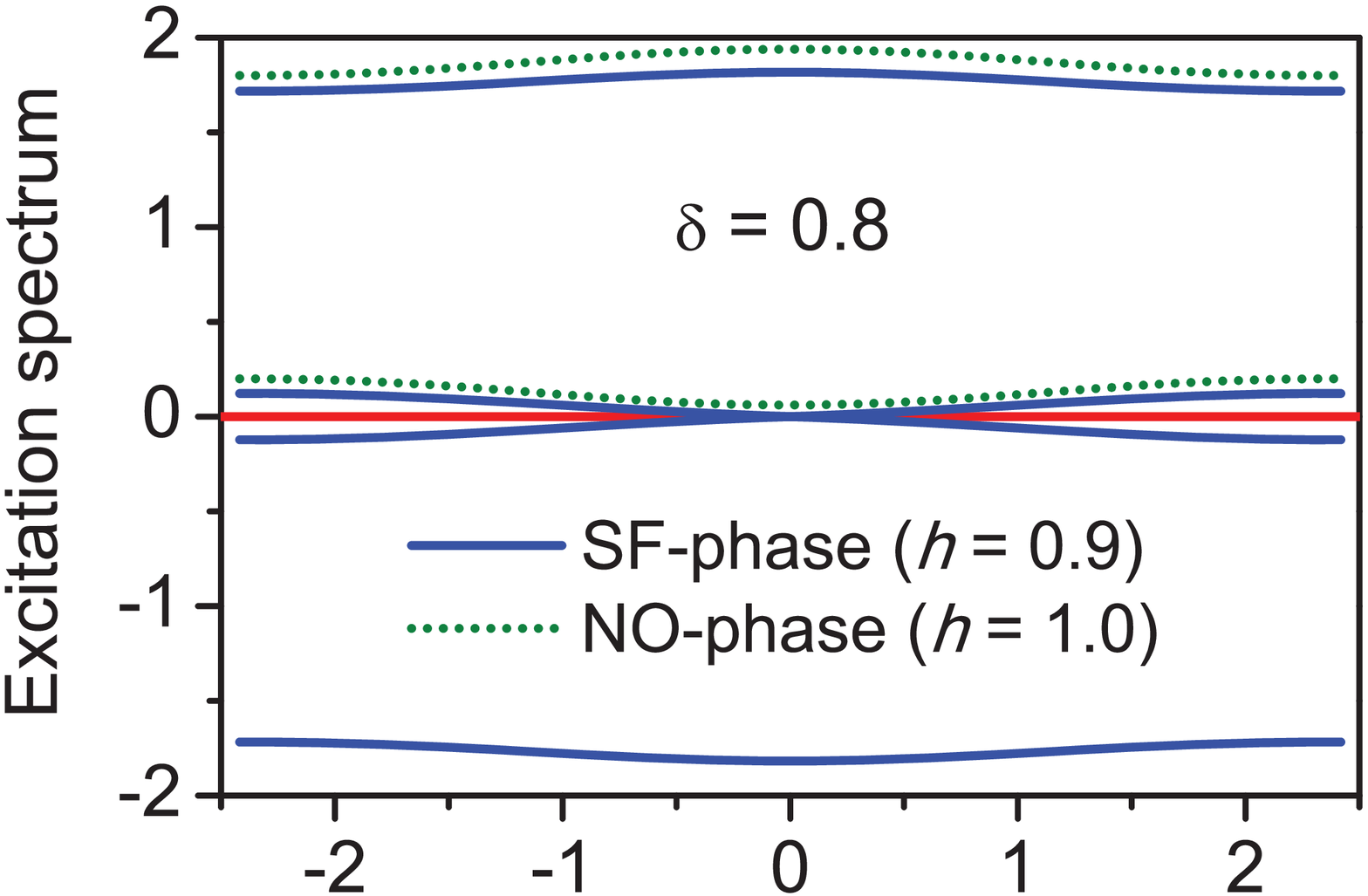}%
\hfill%
\strut%
\vskip-3mm \caption{Forms of spectral branches and their
correspondence to the NO (dotted curves) and SF (solid curves)
phases in the cases where the chemical potential in the NO phase is
located between the bands (left panel; $h=0.6$ in the NO phase and
0.7 in the SF one) and under them (right panel; $h=0.9$ in the SF
phase and 1.0 in the NO one).\,\,The other parameters are
$\delta=0.8$ and {$\Theta=0.05$}.\,\,Hereafter, the energy is
reckoned in the $J(0)$-units from
the chemical potential}%
\label{fig02}\vskip4mm
\end{figure*}

\begin{figure*}
\strut%
\hfill%
\includegraphics[width=5.5cm]{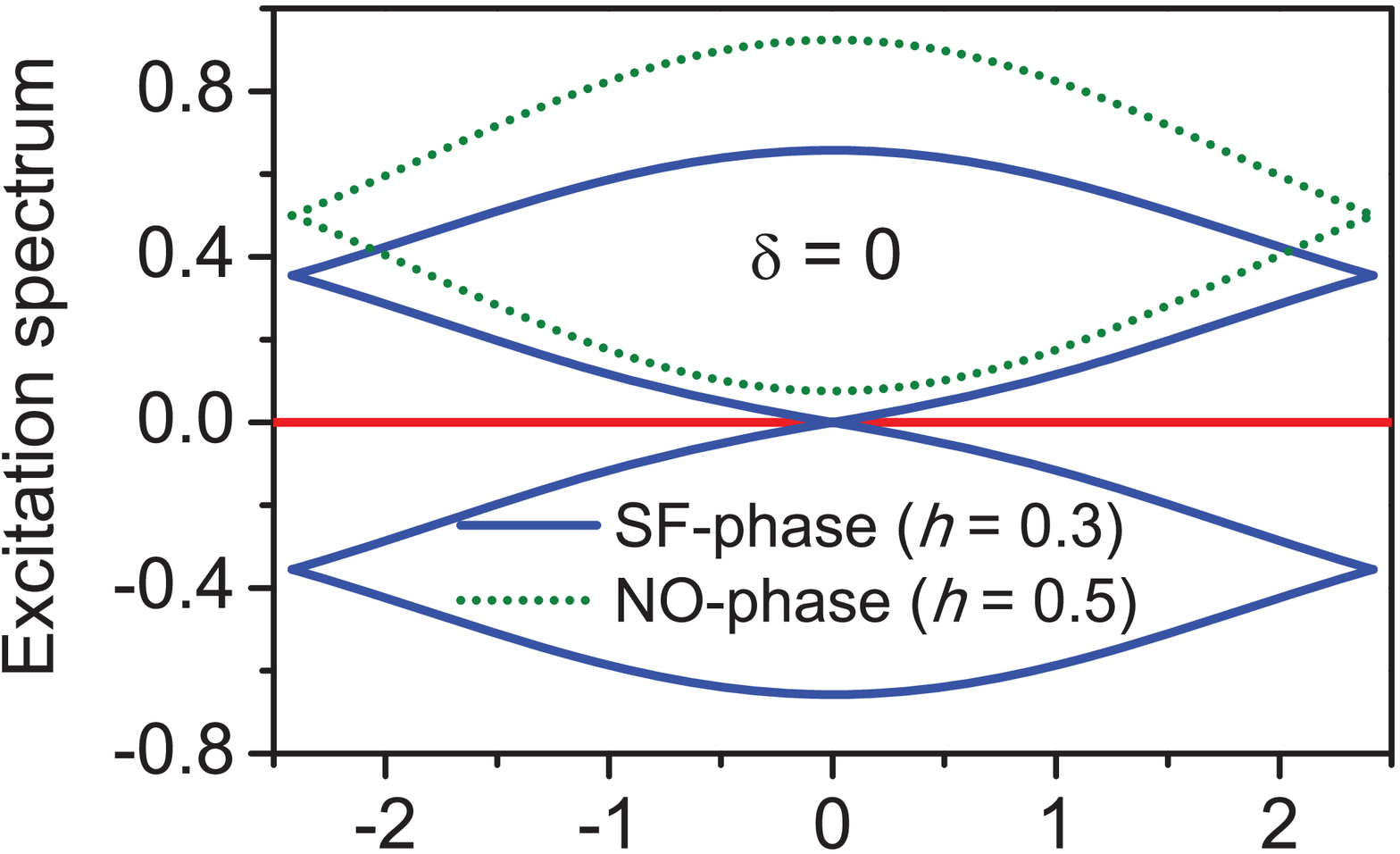}%
\hfill%
\includegraphics[width=5.5cm]{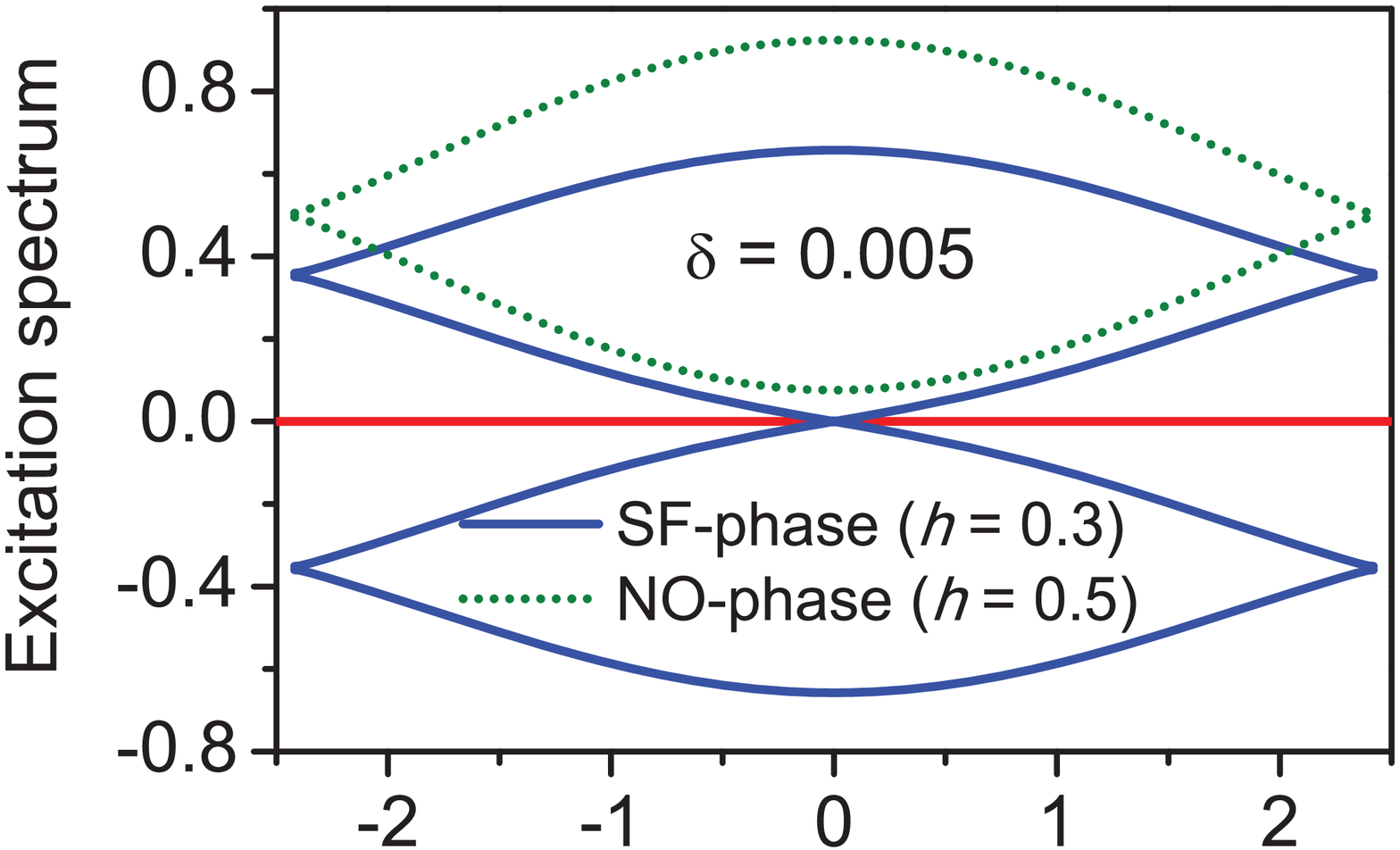}%
\hfill%
\strut%
\\[2ex]%
\strut%
\hfill%
\includegraphics[width=5.5cm]{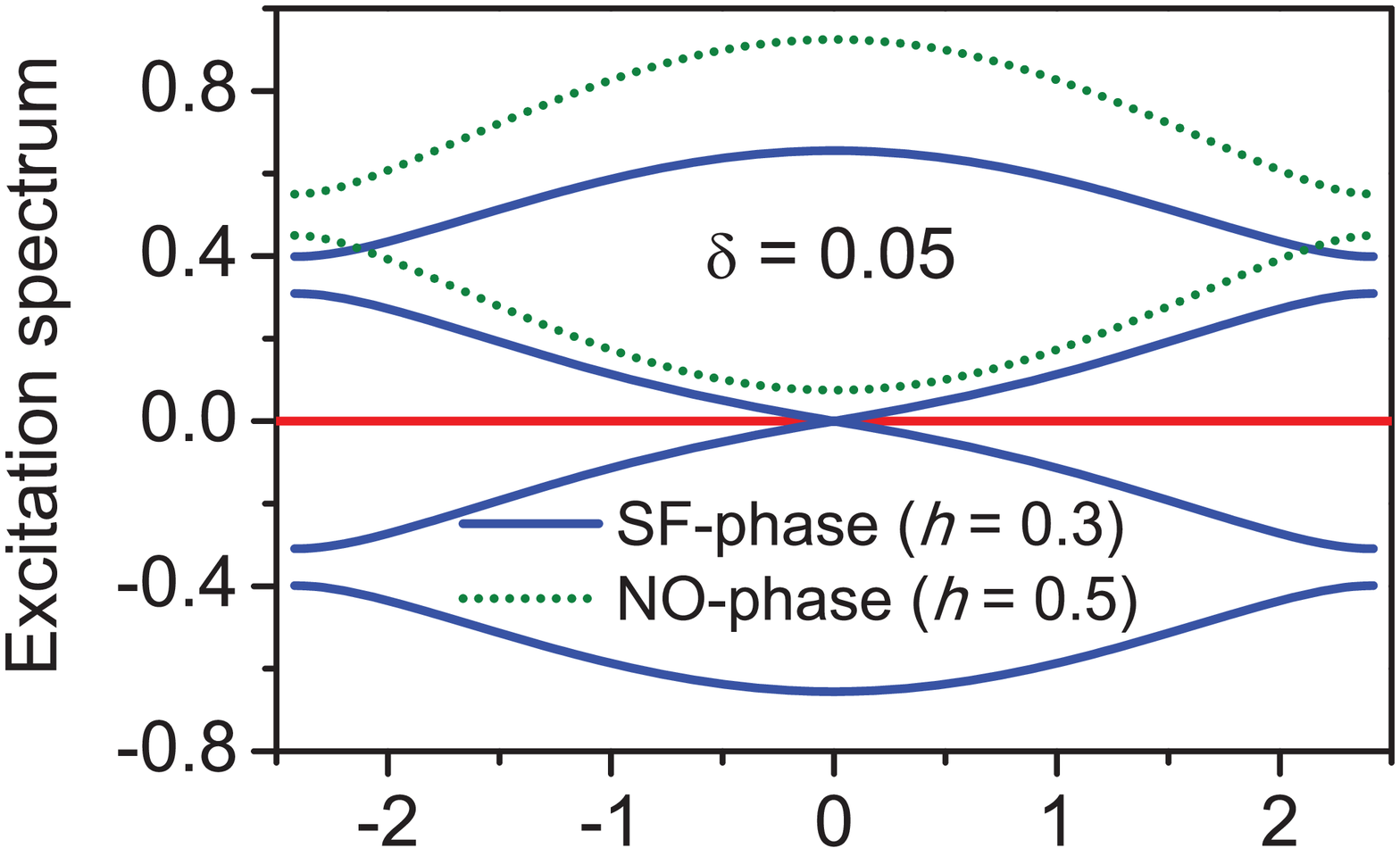}%
\hfill%
\includegraphics[width=5.5cm]{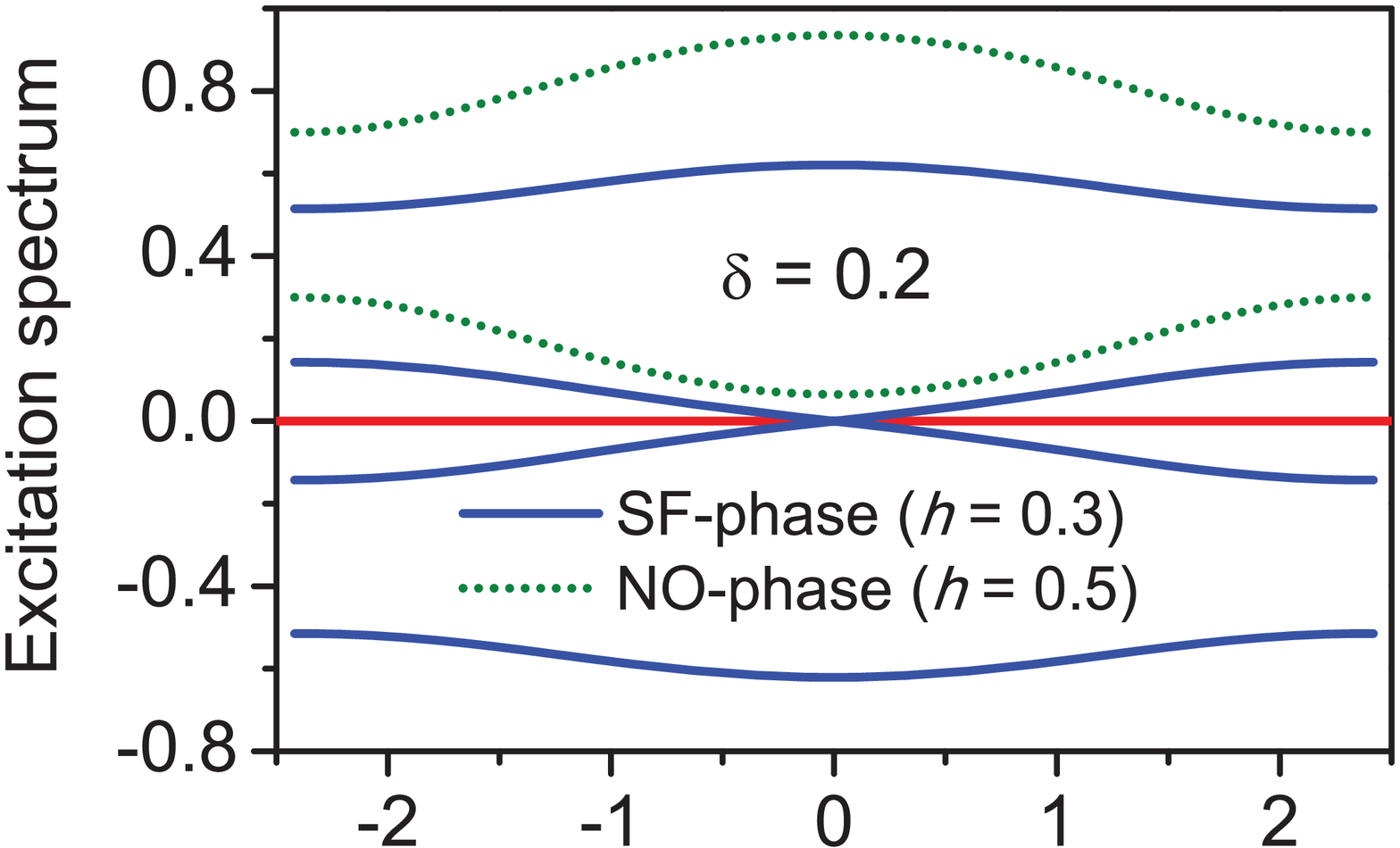}%
\hfill%
\strut%
\vskip-3mm\caption{Appearance of new Dirac points at $\delta=0$, the
divergency of branches at the Brillouin zone boundaries as the
$\delta$ parameter grows, and a change in the behaviour of branches
at the center of the Brillouin zone at the NO\,$\to$\,SF transition
($h=0.3$ in the SF phase and 0.5 in the NO one; $\Theta=0.2$)
}%
\label{fig03}\vspace*{-2mm}
\end{figure*}

The excitation spectrum of bosons in the SF phase consists of four, symmetric
in pairs, branches,
\begin{equation}
\begin{array}{l}
      \displaystyle
    \varepsilon_{1,2}^{\mathrm{(SF)}}(\mathbf{q})
    =
    \pm\left(P_{q}+Q_{q}\right)^{1/2}\!\!,\\[3mm]
   \displaystyle \varepsilon_{3,4}^{\mathrm{(SF)}}(\mathbf{q})
    =
    \pm\left(P_{q}-Q_{q}\right)^{1/2}\!\!,
    \label{Pr-ista2.14}
\end{array}
\end{equation}
where\vspace*{-2mm}
\begin{equation}
\begin{array}{l}
      \displaystyle
    P_{q}
    =
    \frac12\left(E_{\mathrm{A}}^{2}+E_{\mathrm{B}}^{2}\right)+M_{q},
   \\[3mm]
    \displaystyle Q_{q}
    =
    \biggl[
        \frac{1}{4} \left(E_{\mathrm{A}}^{2}-E_{\mathrm{B}}^{2}\right)^{2}
        +
        2N_{q}E_{\mathrm{A}}E_{\mathrm{B}}\,+
      \\[3mm]
           \vphantom{\frac{1}{4}}
        \displaystyle +
        M_{q}\left(E_{\mathrm{A}}^{2}+E_{\mathrm{B}}^{2}\right)\!
    \biggr]^{1/2}\!.
   \label{Pr-ista2.15}
\end{array}
\end{equation}

In comparison with the NO phase, where \cite{a1}
\begin{equation}
\varepsilon_{1,2}(\mathbf{q})=h\pm\sqrt{\delta^{2}+\left\langle \sigma
_{\mathrm{A}}^{z}\right\rangle \left\langle \sigma_{\mathrm{B}}^{z}%
\right\rangle |J(\mathbf{q})|^{2}}, \label{Pr-ista2.16}%
\end{equation}
the number of branches is twice as much.\,\,The duplication of the
specular reflection type with respect to the chemical potential
level takes place at the phase transition, when the variation of
model parameters ($\mu$, $\delta$, or $T$) gives rise to the
situation where the edge of either subband (\ref{Pr-ista2.16})
touches the $\mu$-level.\,\,This situation is illustrated in Figs.~2
and 3 in the cases where the chemical potential is initially (in the
NO phase) located between the subbands or under them.\,\,In the
latter case when the gap in the NO phase spectrum is absent, i.e. at
$\delta=0$, there are the Dirac points with a linear dispersion law
on the Brillouin zone boundary.\,\,They are an analog of such points
in the electron band spectrum of graphene (see, e.g., work
\cite{a2}).\,\,The number of Dirac points doubles after the
transition into the SF phase.\,\,New points of this type appear in
the region of negative energies, if the bosonic band in the NO phase
is located above the chemical potential level, or in the region of
positive energies, if below it (see Fig.~3).

Note that, if the chemical potential $\mu$ is located between the
subbands (in this case, it must be $\delta\neq0$), the spectrum is
reconstructed, and the Dirac points disappear.\,\,The situation
where the chemical potential is located in a close vicinity of the
Dirac points or is imposed onto them is not realized in the case of
hard-core bosons (except for the regions near the critical points,
which are the maximum points for the curves of phase equilibrium on
the phase diagram in Fig.~1).

In the SF phase, two central subbands always touch each other at the
point $\mathbf{q}=0$ at the chemical potential level.\,\,In a
vicinity of this point, the dispersion laws in the subbands are
linear, and, in this sense, they correspond to excitations of the
Bogolyubov type in the interacting Bose gas, which exist in the
presence of a BE condensate \cite{a3}.\vspace*{-2mm}

\begin{figure*}
\strut%
\hfill%
\includegraphics[scale=0.3]{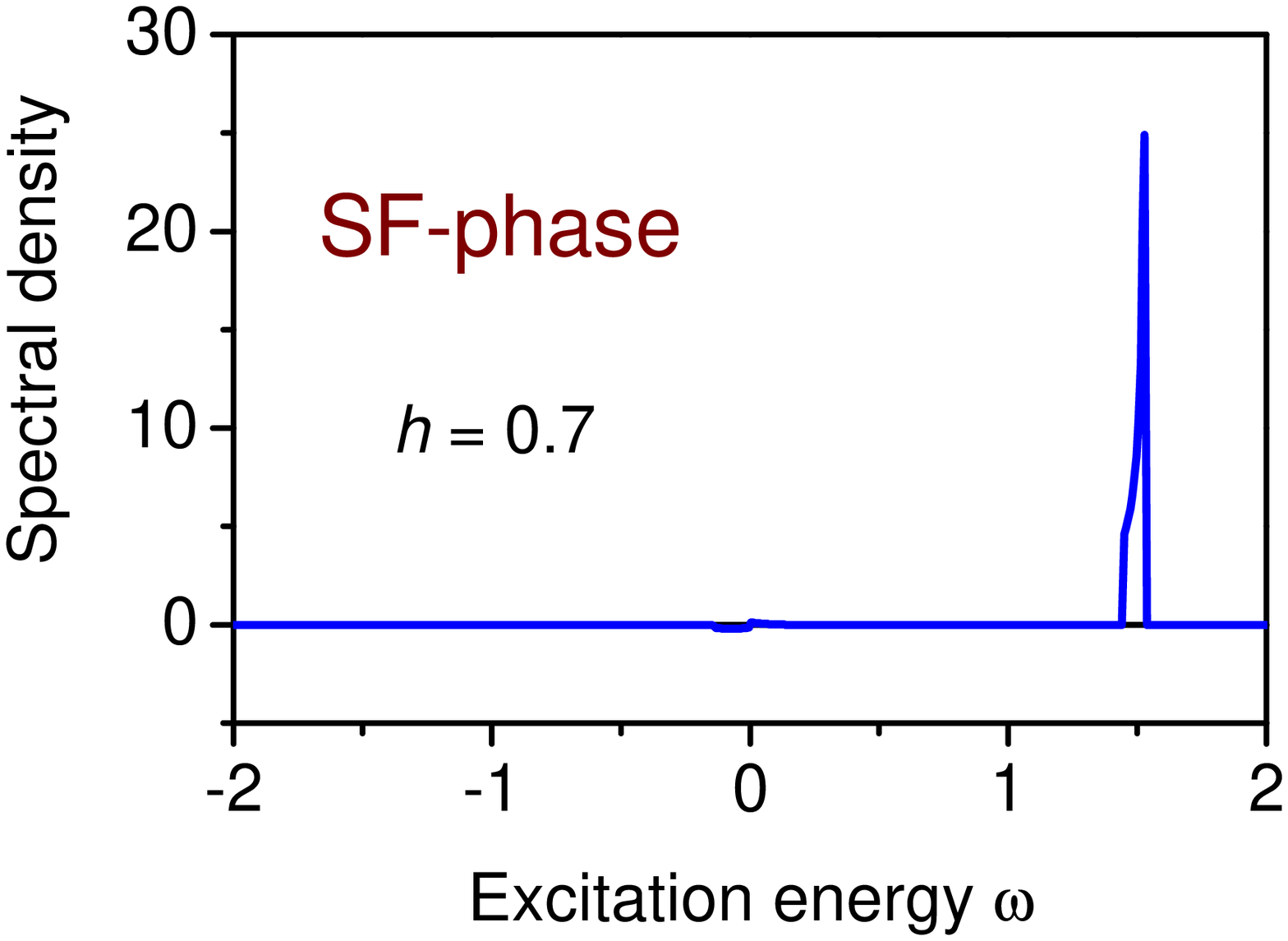}%
\hfill%
\includegraphics[scale=0.3]{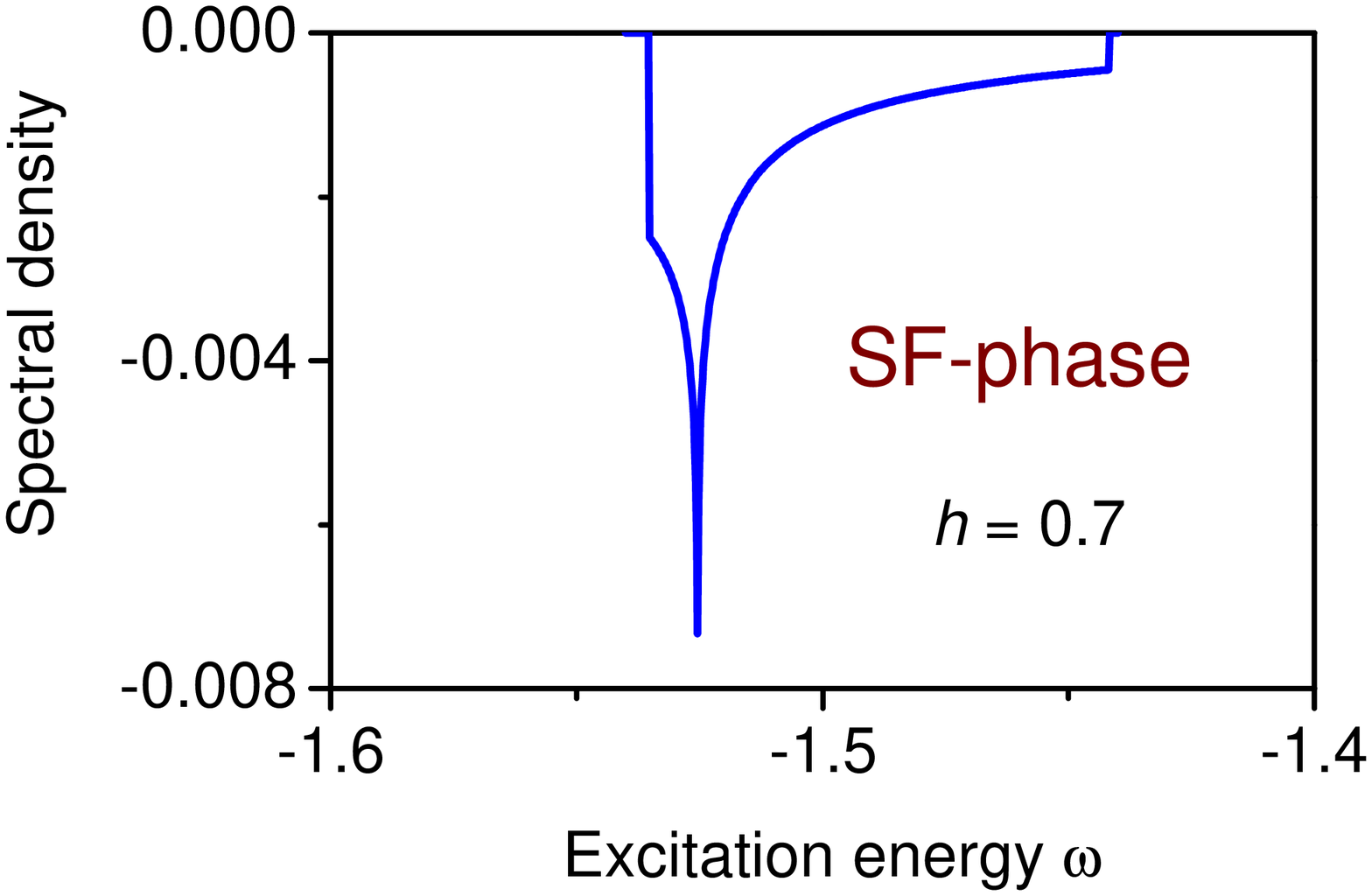}%
\hfill%
\strut%
\\[1.4ex]%
\strut%
\hfill%
\includegraphics[scale=0.3]{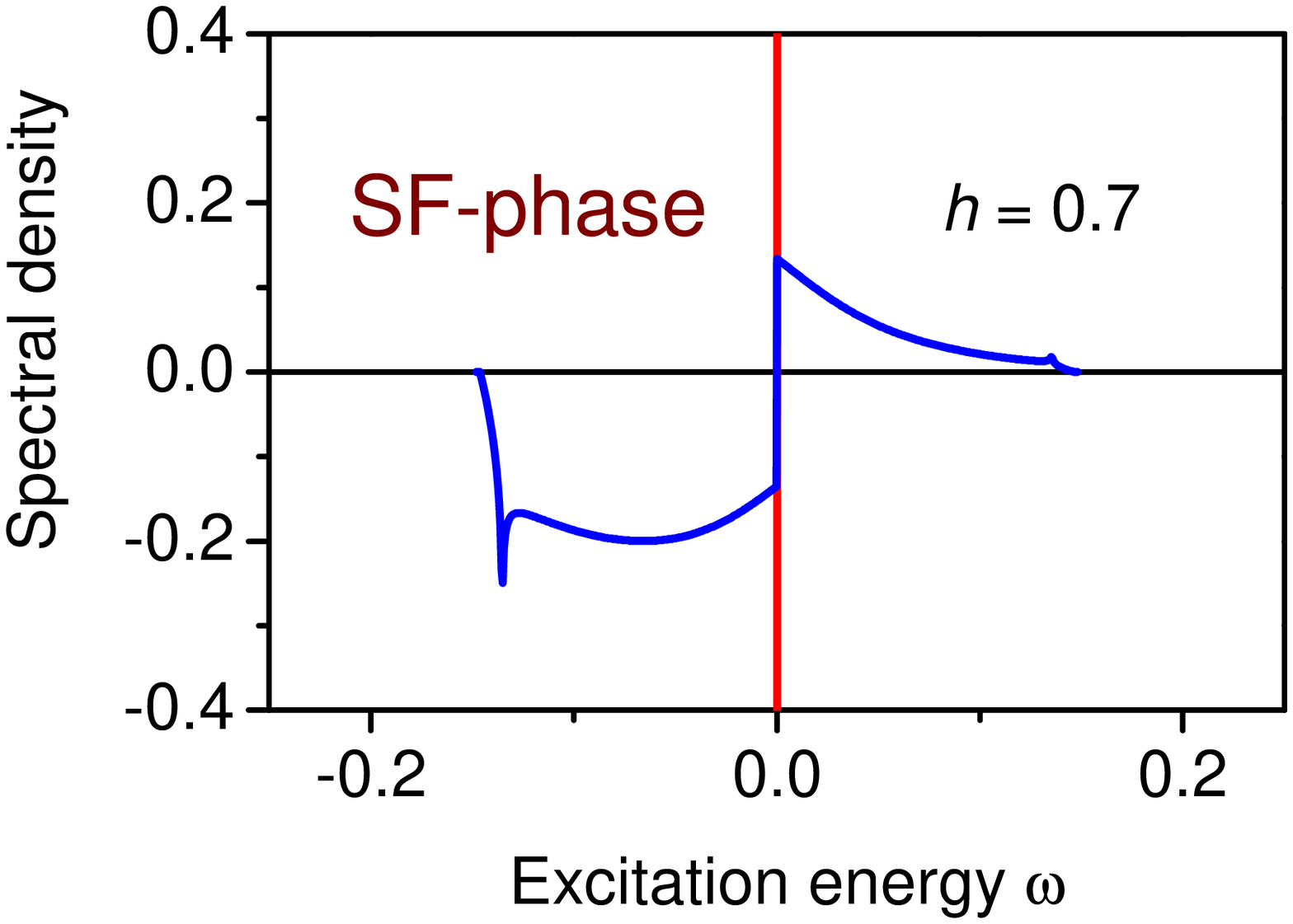}%
\hfill%
\includegraphics[scale=0.3]{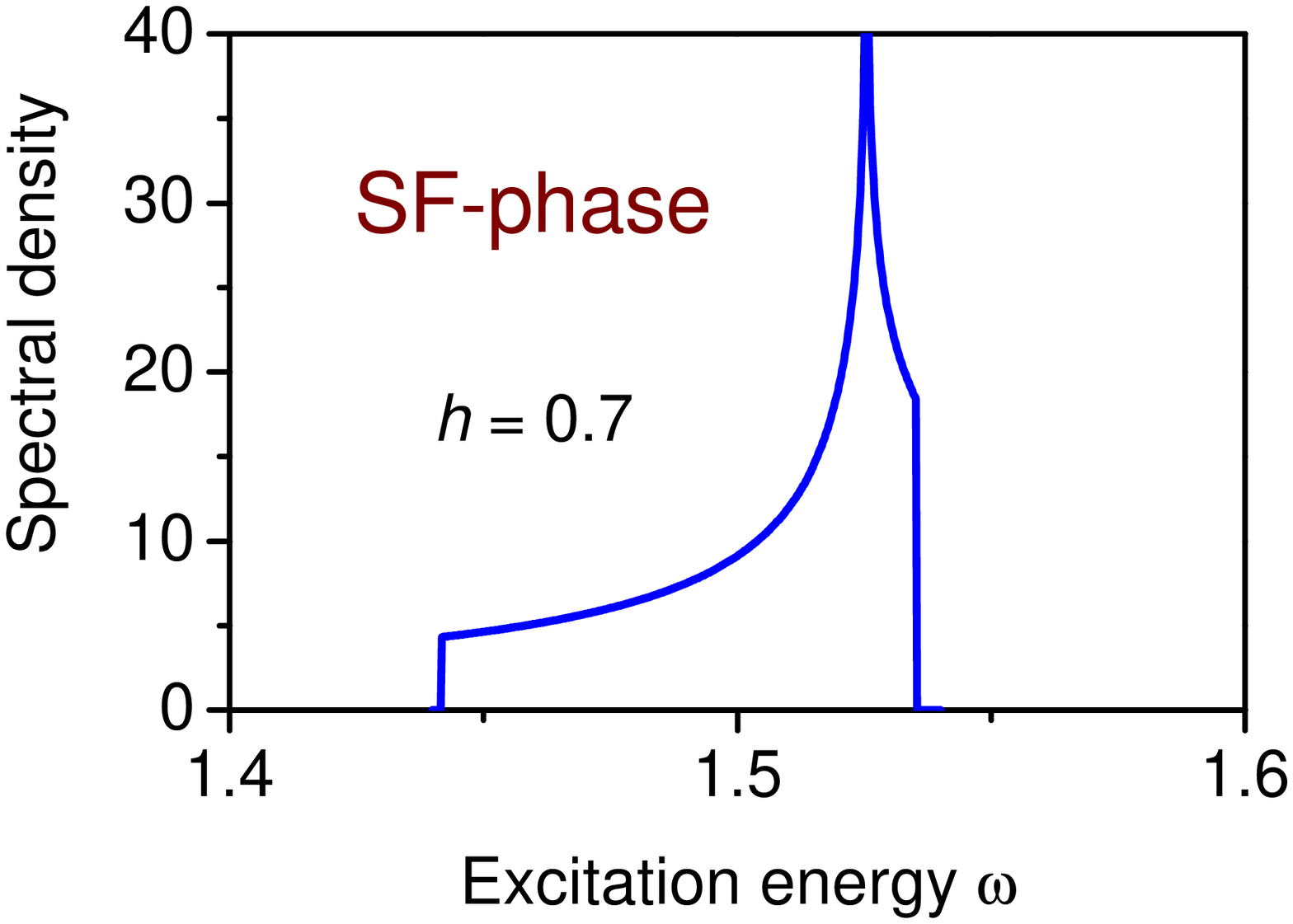}%
\hfill%
\strut%
\vskip-3mm\caption{Spectral density {$\rho_{A}(\omega)$} in the SF
phase: general view (upper left panel) and scaled-up images for
three intervals of the excitation energy {$\hbar\omega$}, where the
spectral density differs from zero (other panels; the weights of
different subbands differ by orders of magnitude).\,\,The
parameter values $J(0)=1$, $\delta=0.8$, $h=0.7$, and $\Theta=0.05$}%
\label{fig04}
\end{figure*}

\begin{figure*}
\includegraphics[width=0.3\textwidth]{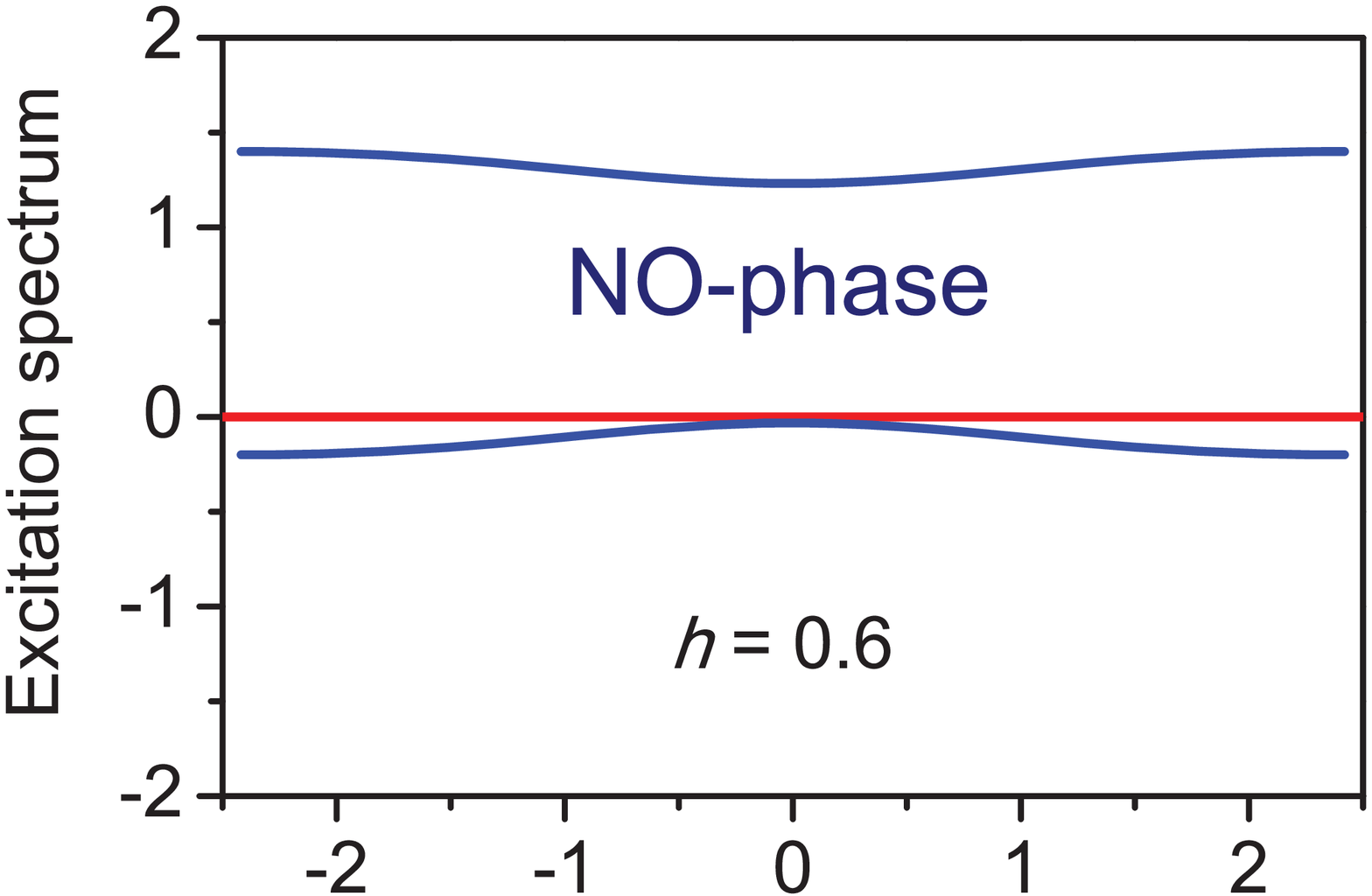}%
\hfill%
\includegraphics[width=0.32\textwidth]{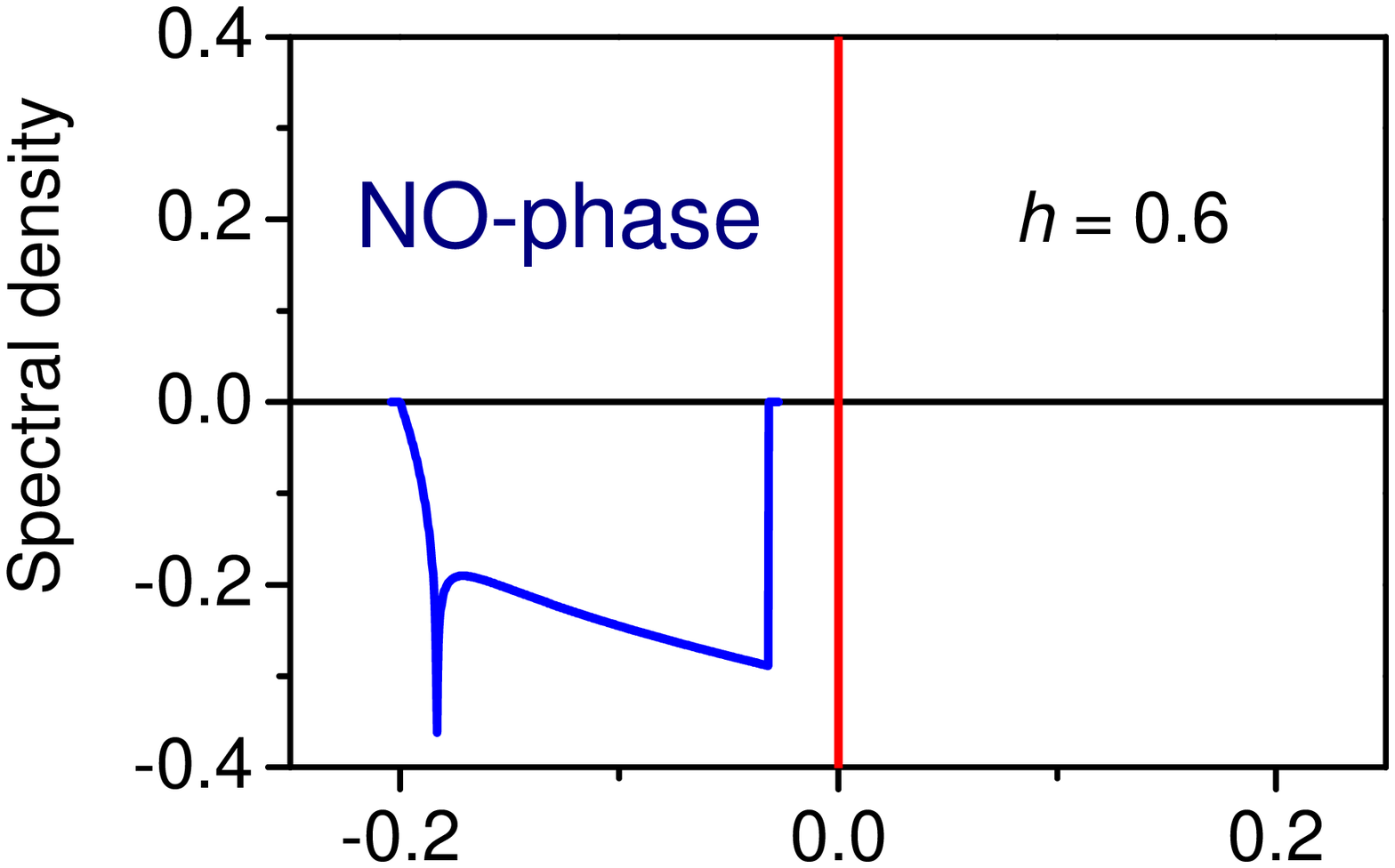}%
\hfill%
\includegraphics[width=0.32\textwidth]{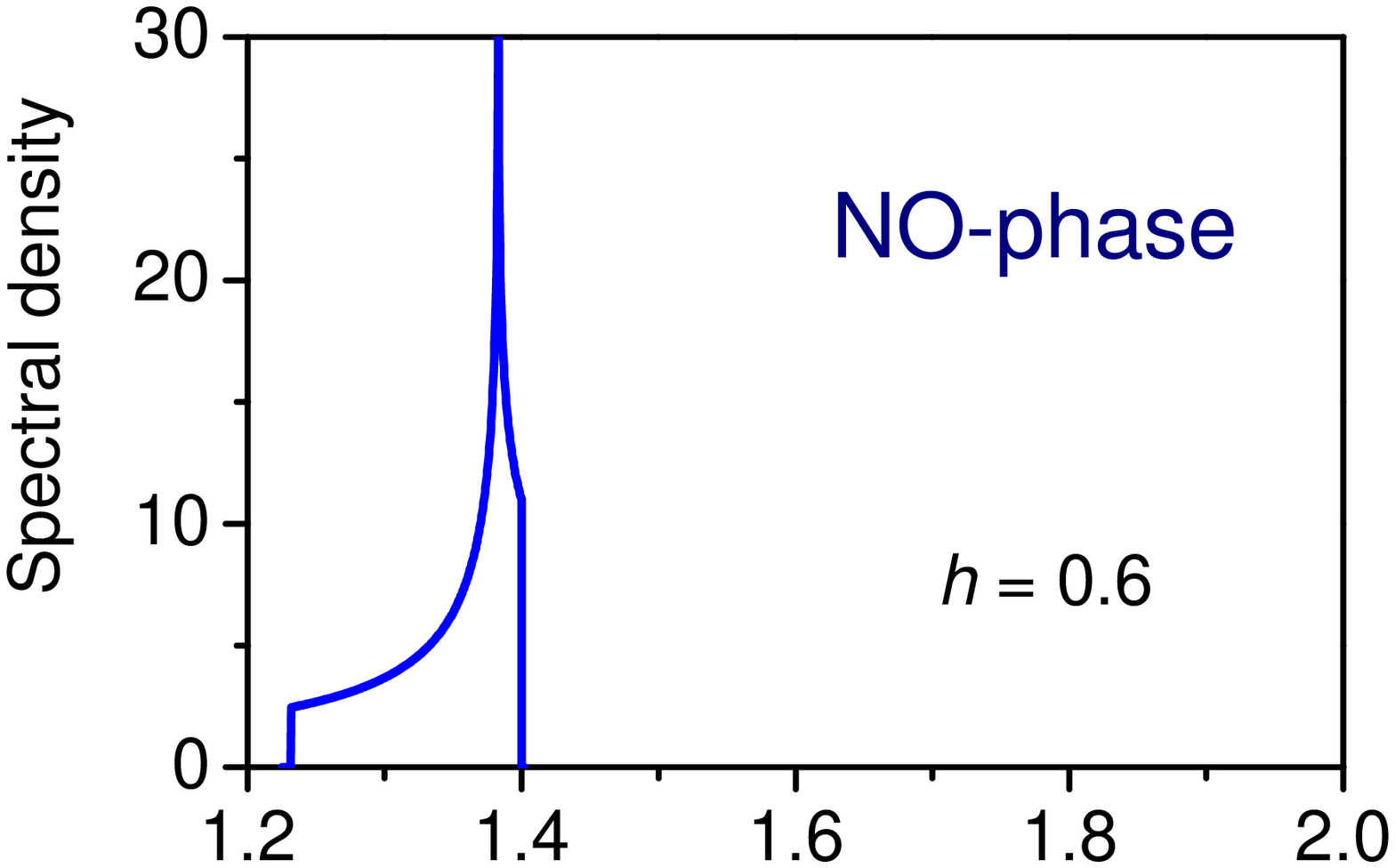}%
\\[1.4ex]%
\includegraphics[width=0.3\textwidth]{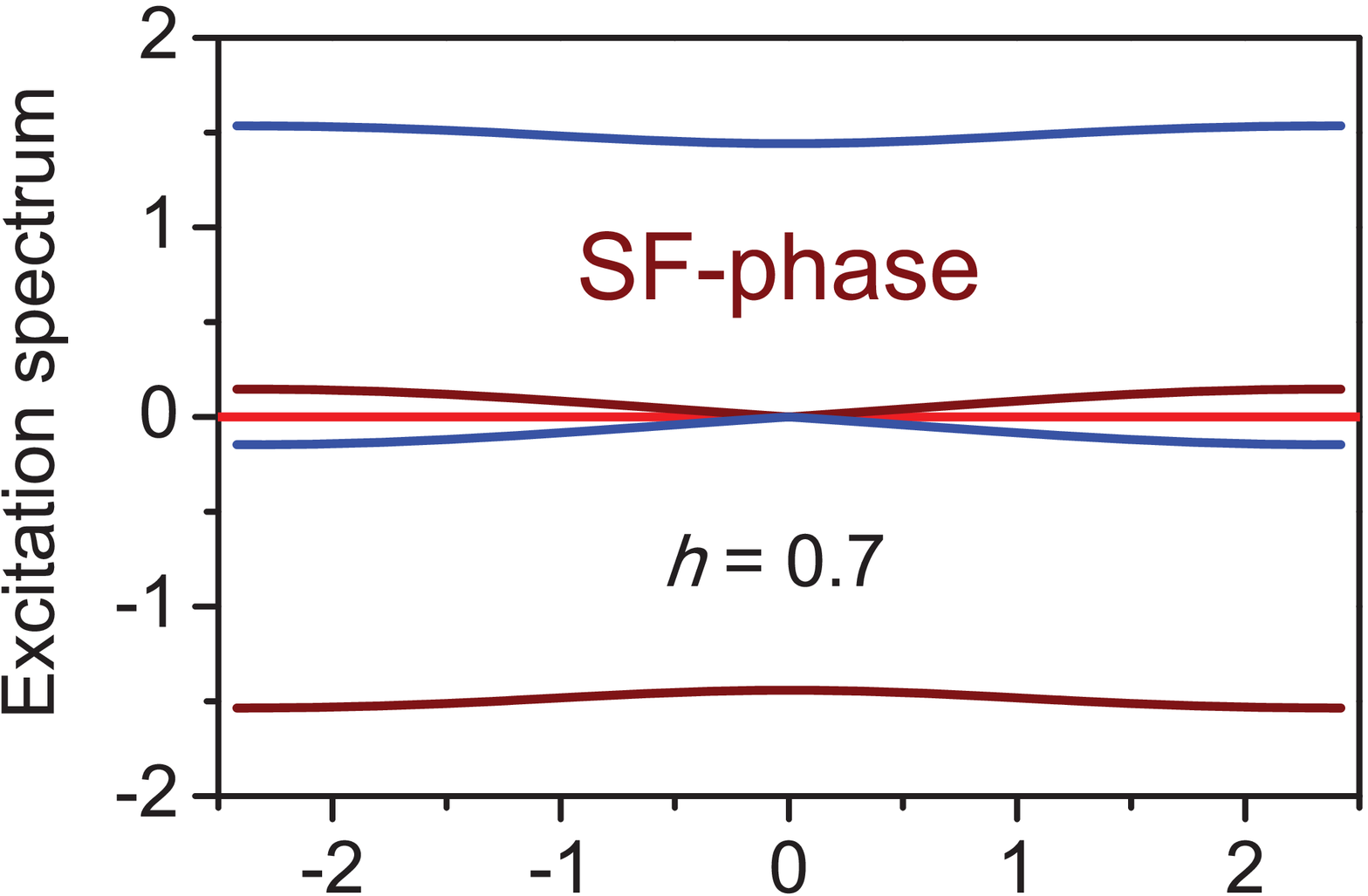}%
\hfill%
\includegraphics[width=0.32\textwidth]{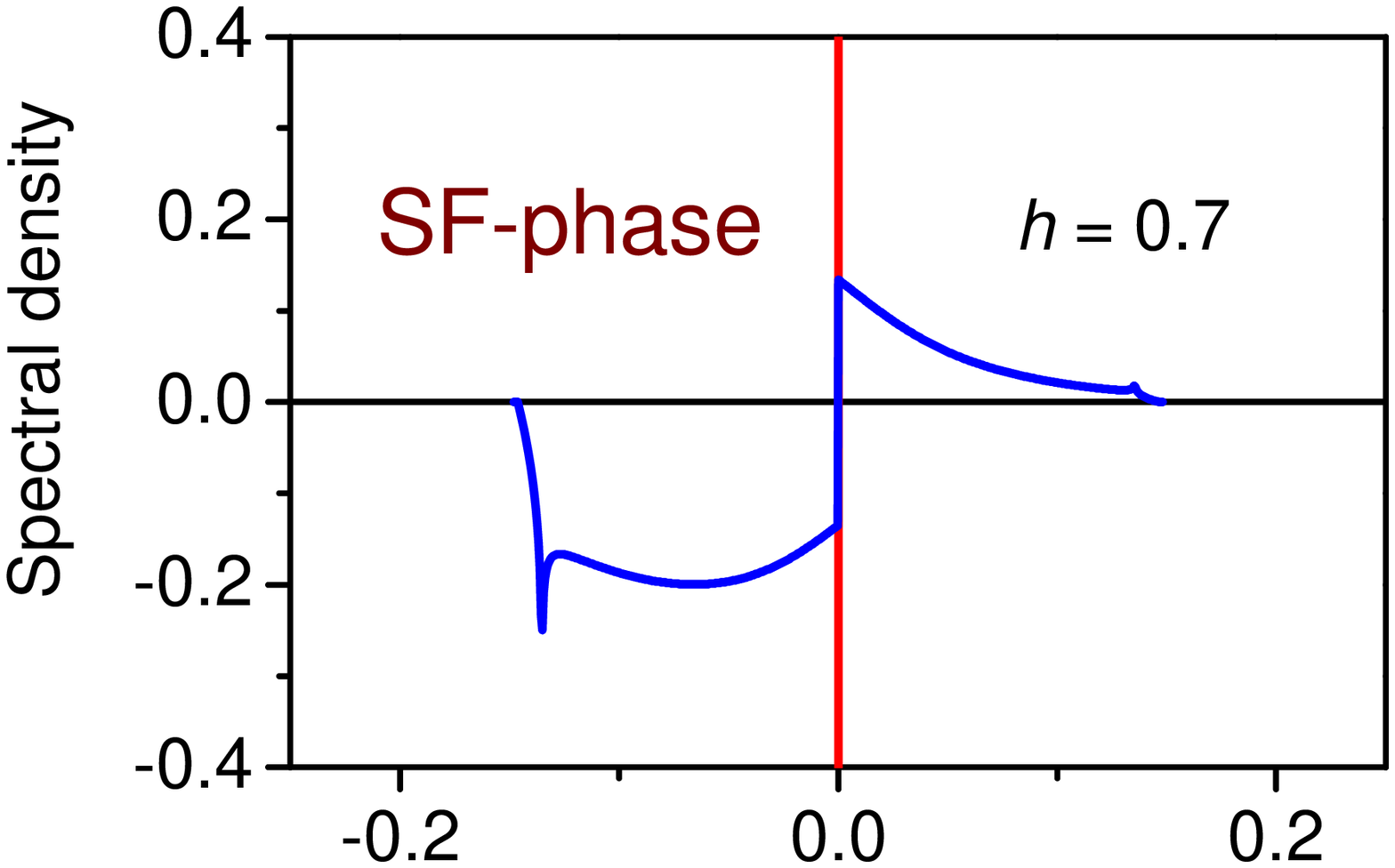}%
\hfill%
\includegraphics[width=0.32\textwidth]{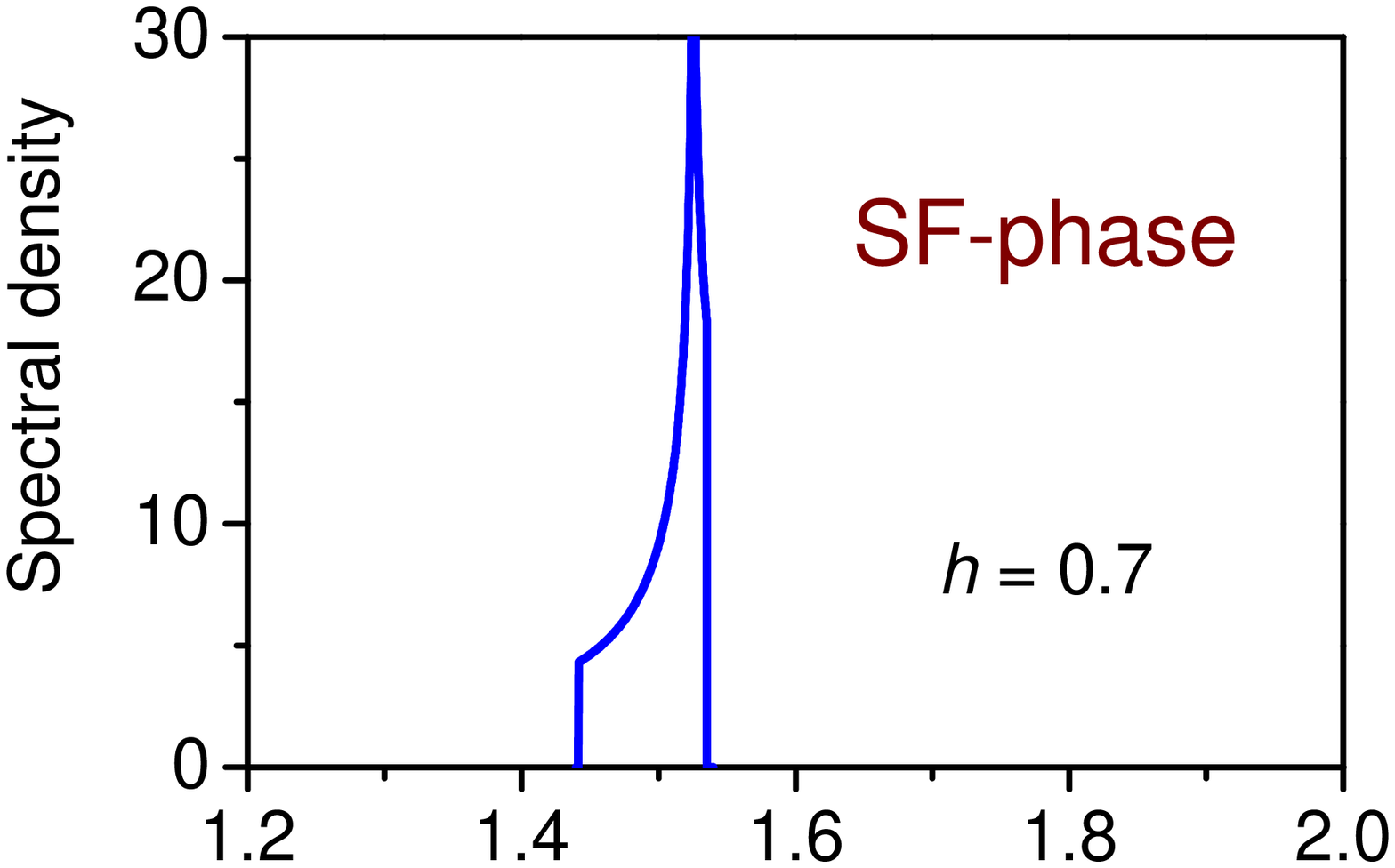}%
\\[1.4ex]%
\includegraphics[width=0.3\textwidth]{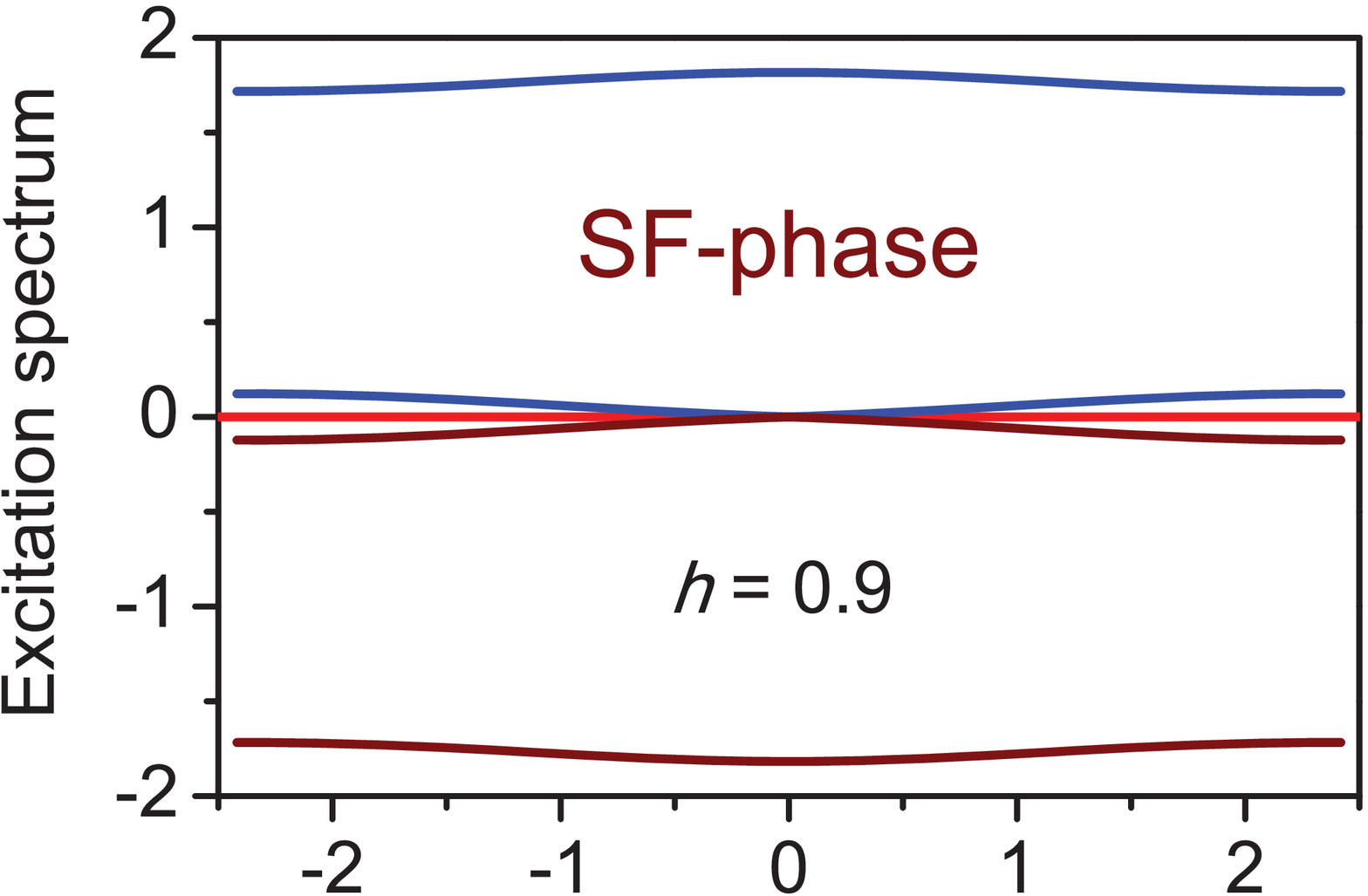}%
\hfill%
\includegraphics[width=0.32\textwidth]{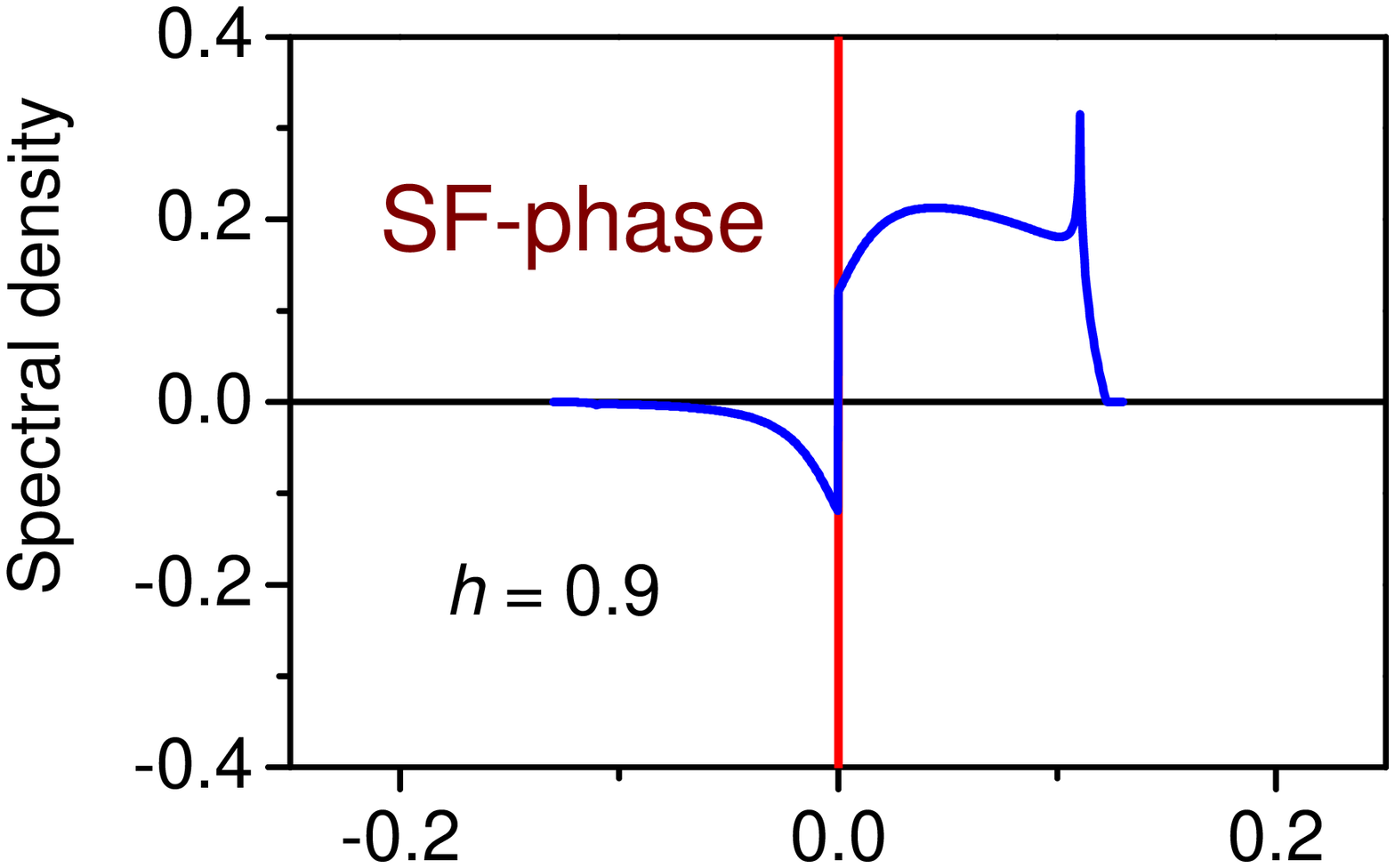}%
\hfill%
\includegraphics[width=0.32\textwidth]{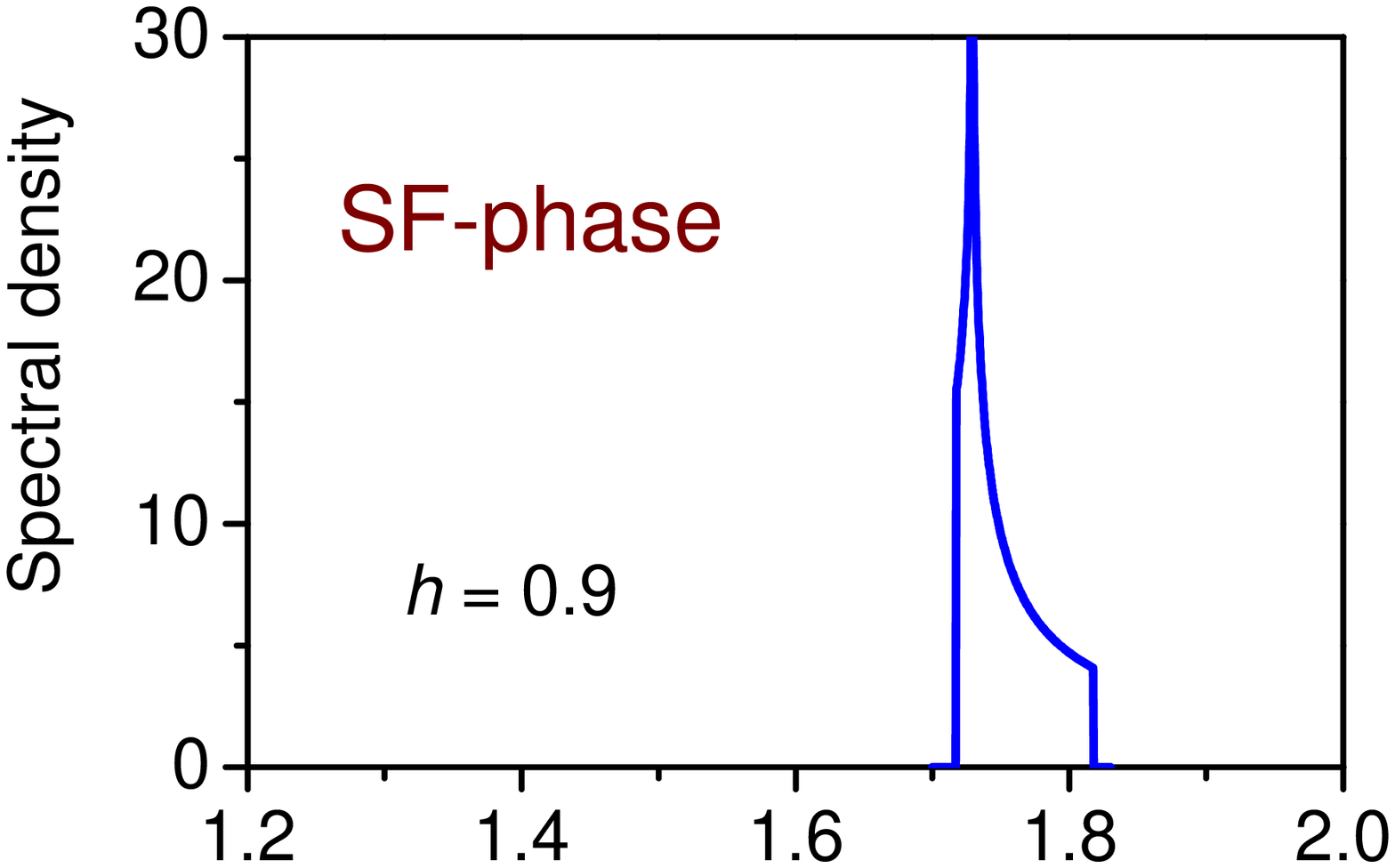}%
\\[1.4ex]%
\includegraphics[width=0.3\textwidth]{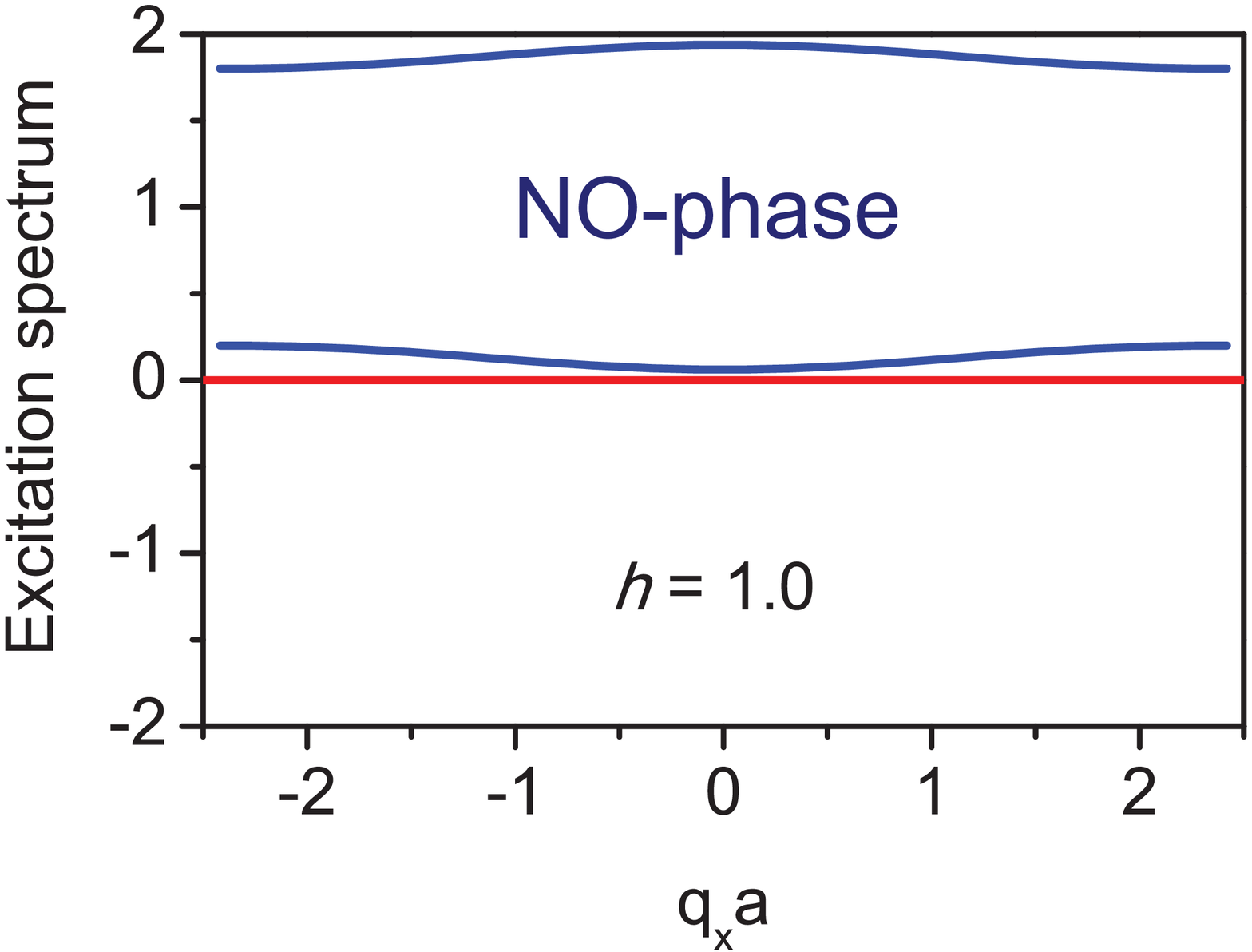}%
\hfill%
\includegraphics[width=0.32\textwidth]{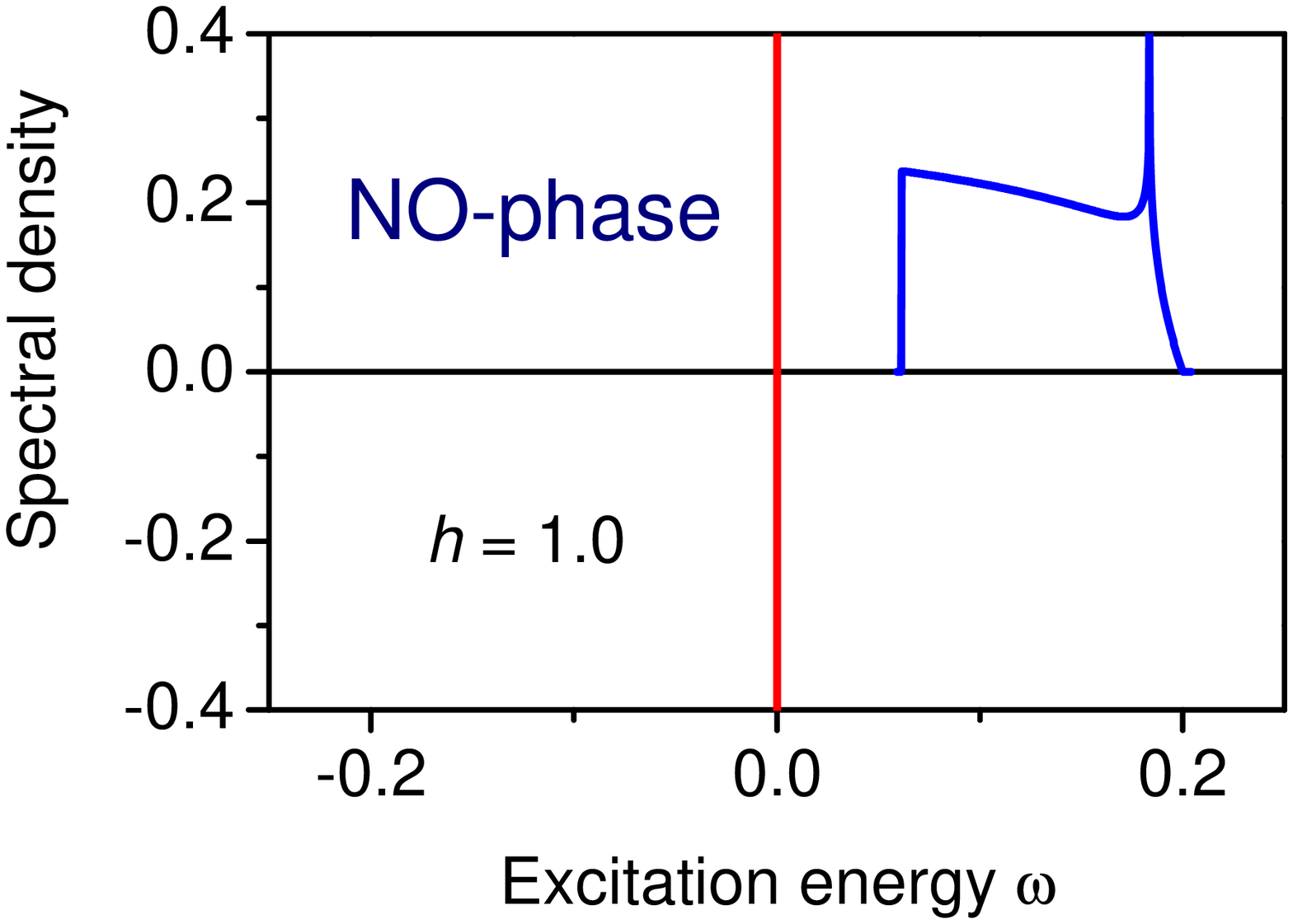}%
\hfill%
\includegraphics[width=0.32\textwidth]{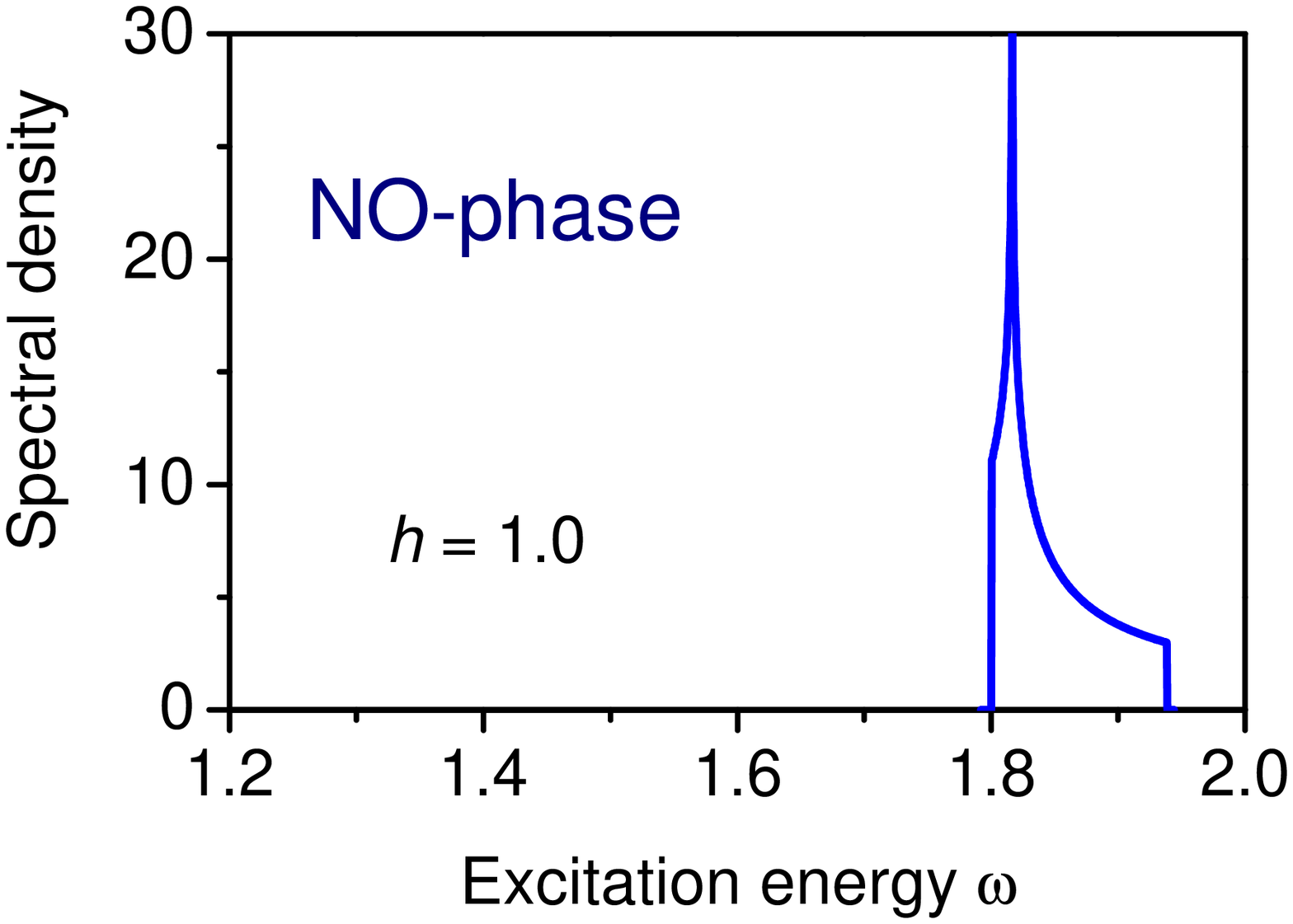}%
\vskip-3mm\caption{Spectrum (left panels) and spectral density
{$\rho_{A}(\omega)$} (central and right panels) in the NO and SF
phases at various energies $h$.\,\,The other parameters are
$J(0)=1$, $\delta=0.8$, and $\Theta=0.05$. The plots of
$\rho_A(\omega)$ for the lowest branch in the SF phase are not shown
}%
\label{fig05}\vspace*{-2mm}
\end{figure*}

\section{Spectral\\ Density \boldmath$\rho_{\alpha}(\omega)$ in the SF Phase}

The spectral density function for single particle bosonic
excitations calculated per one lattice site will be designated here
as
\[
    \rho_\alpha (\omega)
    =
    -\frac{2}{N}\sum_{\mathbf q}
    \mathrm{Im}\, \langle\langle b_\alpha|b_\alpha^+
    \rangle\rangle_{\mathbf q, \omega +
    \mathrm{i}\varepsilon}=
    \]\vspace*{-7mm}
    \begin{equation}
    =
    -\frac{2}{N}\sum_{\mathbf q}
    \mathrm{Im}\, \langle\langle S_\alpha^{+}|S_\alpha^-
    \rangle\rangle_{\mathbf q, \omega +
    \mathrm{i}\varepsilon}.
    \label{Pr-ista3.1}
\end{equation}
Proceeding from formula (\ref{Pr-ista2.10}) and making a decomposition into
simple fractions, the function $\rho_{\alpha}(\omega)$ can be written as
follows:
\begin{equation}
\rho_{\alpha}(\omega)=\frac{1}{N}\sum_{q}\langle\sigma_{\alpha}^{z}\rangle
\sum_{i=1}^{4}A_{i}^{\alpha}(\mathbf{q})\delta\left(\!  \omega-\frac
{\varepsilon_{i}(\mathbf{q})}{\hbar}\!\right)\!, \label{Pr-ista3.2}%
\end{equation}
where\vspace*{-2mm}
\begin{equation}
\begin{array}{l}
      \displaystyle
    A^{\alpha}_{1,2}(\mathbf{q})
    =
    \frac{P_{q}^{\alpha}\left(\hbar\omega=\varepsilon_{1,2}(\mathbf{q})\right)}
        {4Q_{q}\varepsilon_{1,2}(\mathbf{q})},
    \\[5mm]
     \displaystyle A_{3,4}^{\alpha}(\mathbf{q})
    =
    -
    \frac{P_{q}^{\alpha}\left(\hbar\omega=\varepsilon_{3,4}(\mathbf{q})\right)}
        {4Q_{q}\varepsilon_{3,4}(\mathbf{q})}.
    \label{Pr-ista3.3}
\end{array}
\end{equation}
The summation over the wave vectors is carried out in accordance with the formula
\[
    \frac{1}{N} \sum_{\mathbf{q}} \Phi (|J({\mathbf{q}})|^2)
    =
    \frac{1}{N} \sum_{{\mathbf q} } \Phi (t^2|\gamma_{ q}|^2)=
   \]\vspace*{-5mm}
  \begin{equation}
    =
    \int dx \rho_0(x) \Phi (t^2 x),
    \label{Pr-ista3.4}
\end{equation}
where\vspace*{-3mm}
\[
|\gamma_{\mathbf{q}}|^{2}  =1+4\cos\left(\!
q_{x}\frac{a\sqrt{3}}{2}\!\right) \cos\left(\!
q_{y}\frac{3}{2}a\!\right)+
\]\vspace*{-5mm}
\[
+\,4\cos^{2}\left( \! q_{x}\frac{q\sqrt{3}}{2}\!\right)\!\!,
\]
and $\rho_{0}(x)=\frac{1}{N}\sum_{\mathbf{q}}\delta(x-|\gamma_{\mathbf{q}%
}|^{2})$ is an auxiliary function that characterizes the distribution over the
squared energy and is related to the band density of states $g(\mathcal{E})$
for the graphene lattice as follows:
\begin{equation}
\rho_{0}(x)=\frac{1}{2\sqrt{x}}g(\sqrt{x}). \label{Pr-ista3.5a}%
\end{equation}
The function $g(\mathcal{E})$ looks like
\[
g(\mathcal{E})
    =
    \frac{2}{\pi^{2}}
    \frac{|\mathcal{E}|}{\sqrt{Z_{0}}}
    F\left(\!\frac{\pi}{2},\sqrt{\frac{Z_{1}}{Z_{0}}}\!\right)\!\!,
\]

\noindent where
\begin{equation}
\begin{array}{l}
\displaystyle {Z_0} =   \left\{\!\!
                \begin{array}{ll}
                             (1+|\mathcal{E}| )^2 -\frac14 \left(|\mathcal{E}|^2-1\right)^{\!2}\!\!,  &  |\mathcal{E}| \leqslant 1, \\
                                                   4|\mathcal{E}|,  &   1\leqslant  |\mathcal{E}| \leqslant 3,
               \end{array}
               \right.\\[5mm]

       \displaystyle {Z_1} =  \left\{\!\!
                \begin{array}{ll}
                             4|\mathcal{E}|,  &  |\mathcal{E}| \leqslant 1, \\
        (1+|\mathcal{E}|)^2 -\frac14 (|\mathcal{E}|^2-1)^{2},  &  1 \leqslant  |\mathcal{E}| \leqslant 3,
                \end{array}
                \right.
                 \label{Pr-ista3.5}
\end{array}
\end{equation}
and $F(\pi/2,y)$ is the complete elliptic integral of the first kind (see
works \cite{a4,a5}).

\begin{figure*}
\strut%
\hfill%
\includegraphics[height=0.15\textheight]{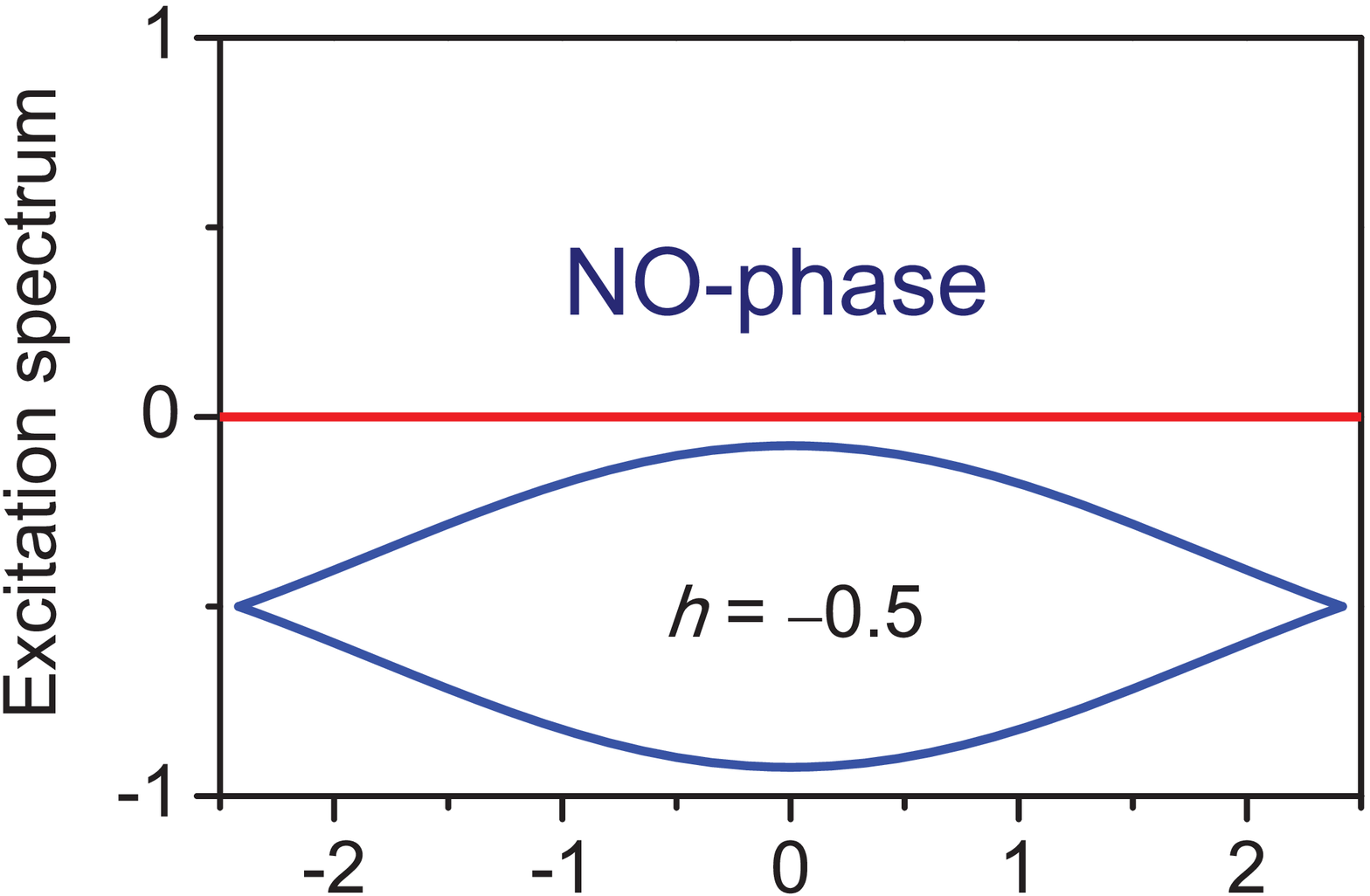}%
\hfill%
\includegraphics[height=0.15\textheight]{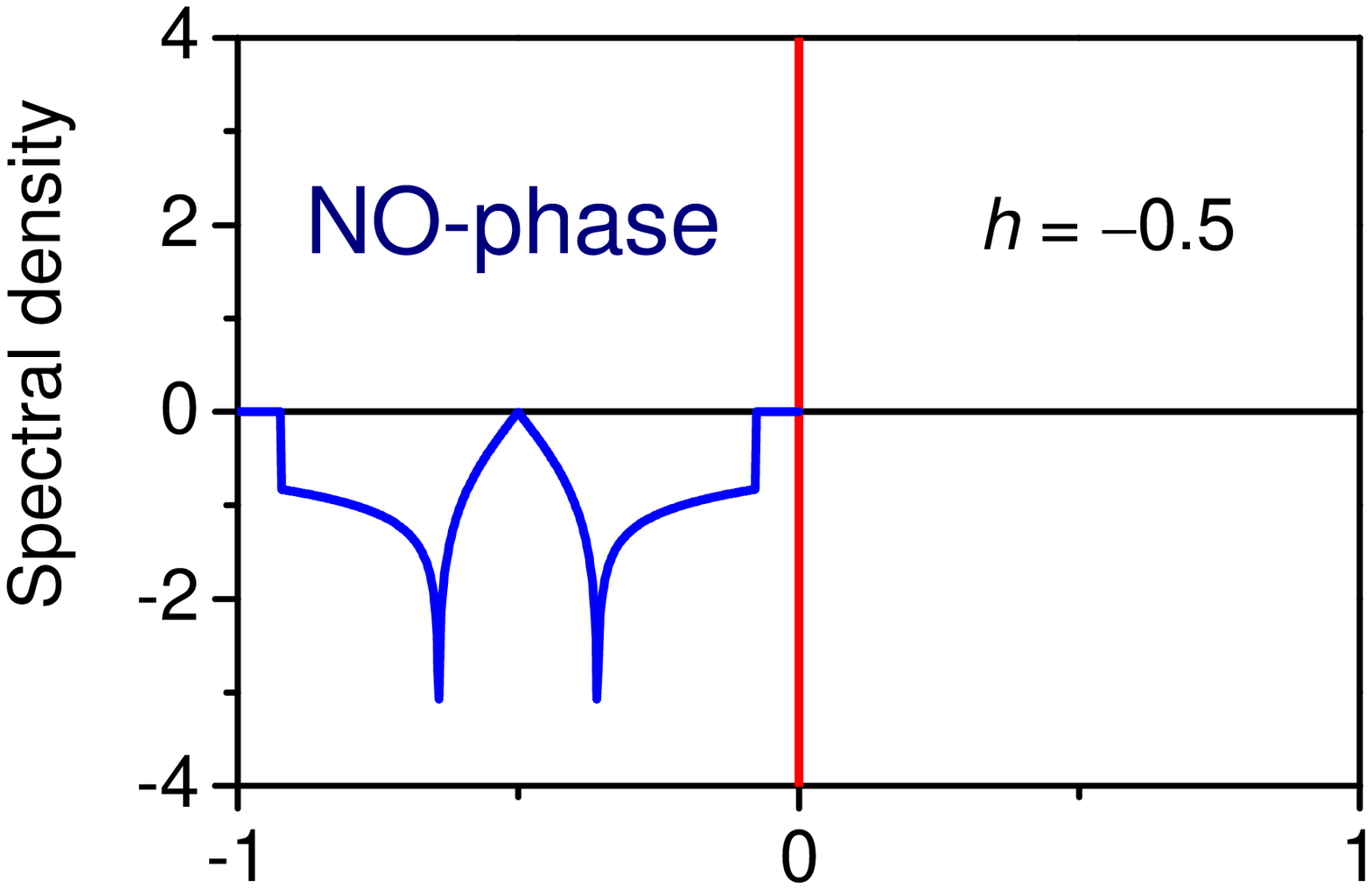}%
\hfill%
\strut%
\\[0.5mm]%
\strut%
\hfill%
\includegraphics[height=0.15\textheight]{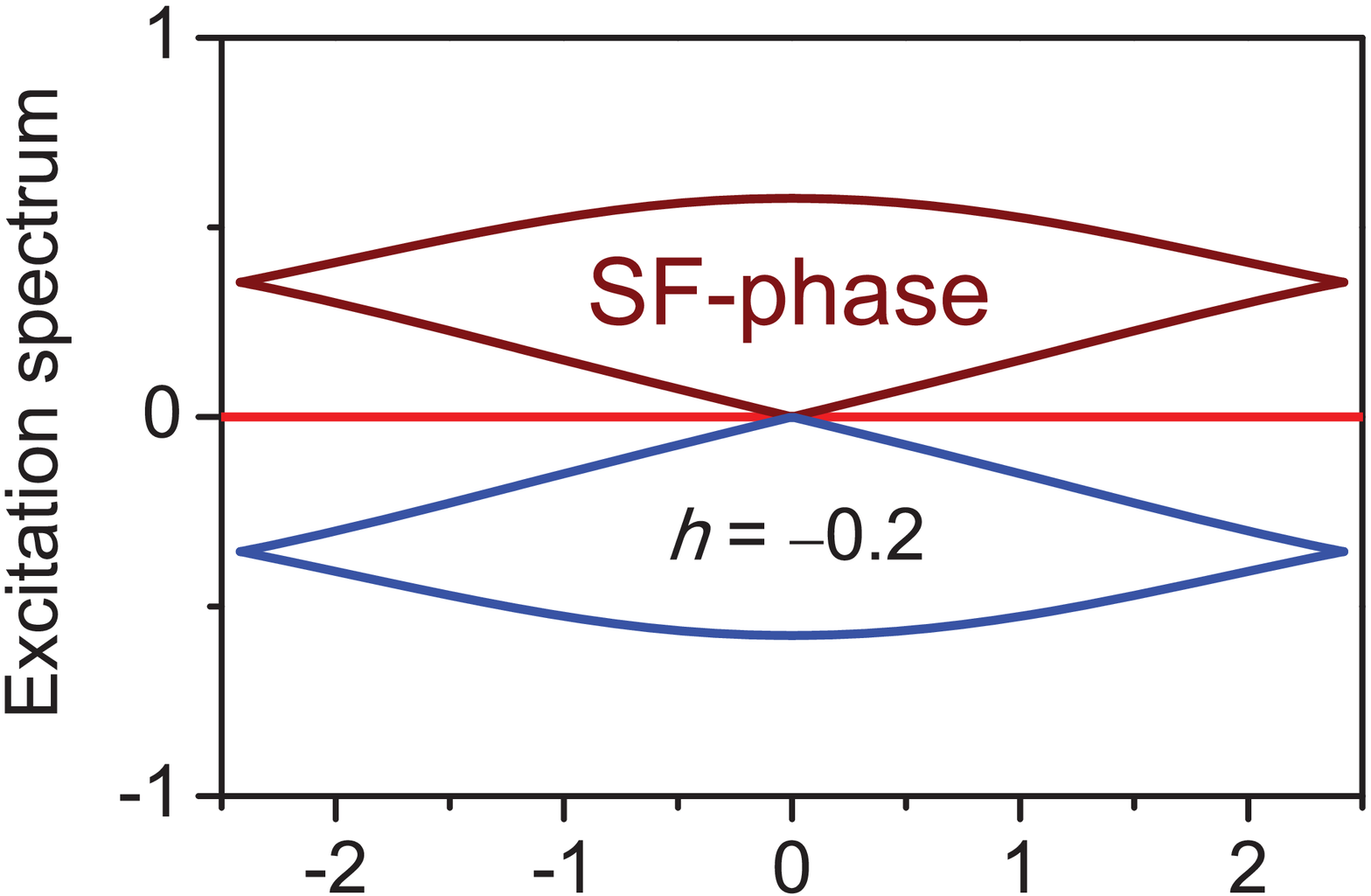}%
\hfill%
\includegraphics[height=0.15\textheight]{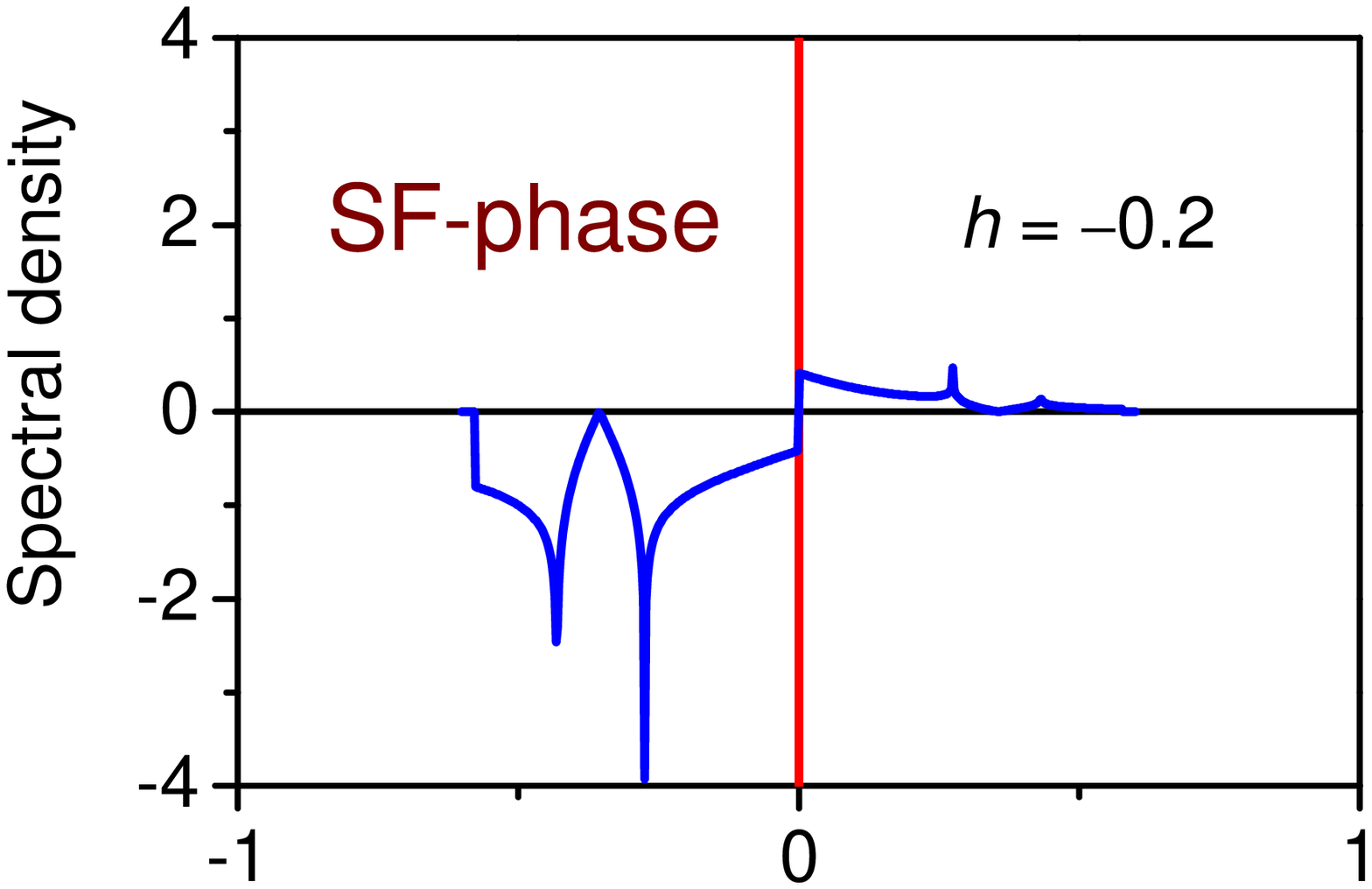}%
\hfill%
\strut%
\\[0.5mm]%
\strut%
\hfill%
\includegraphics[height=0.15\textheight]{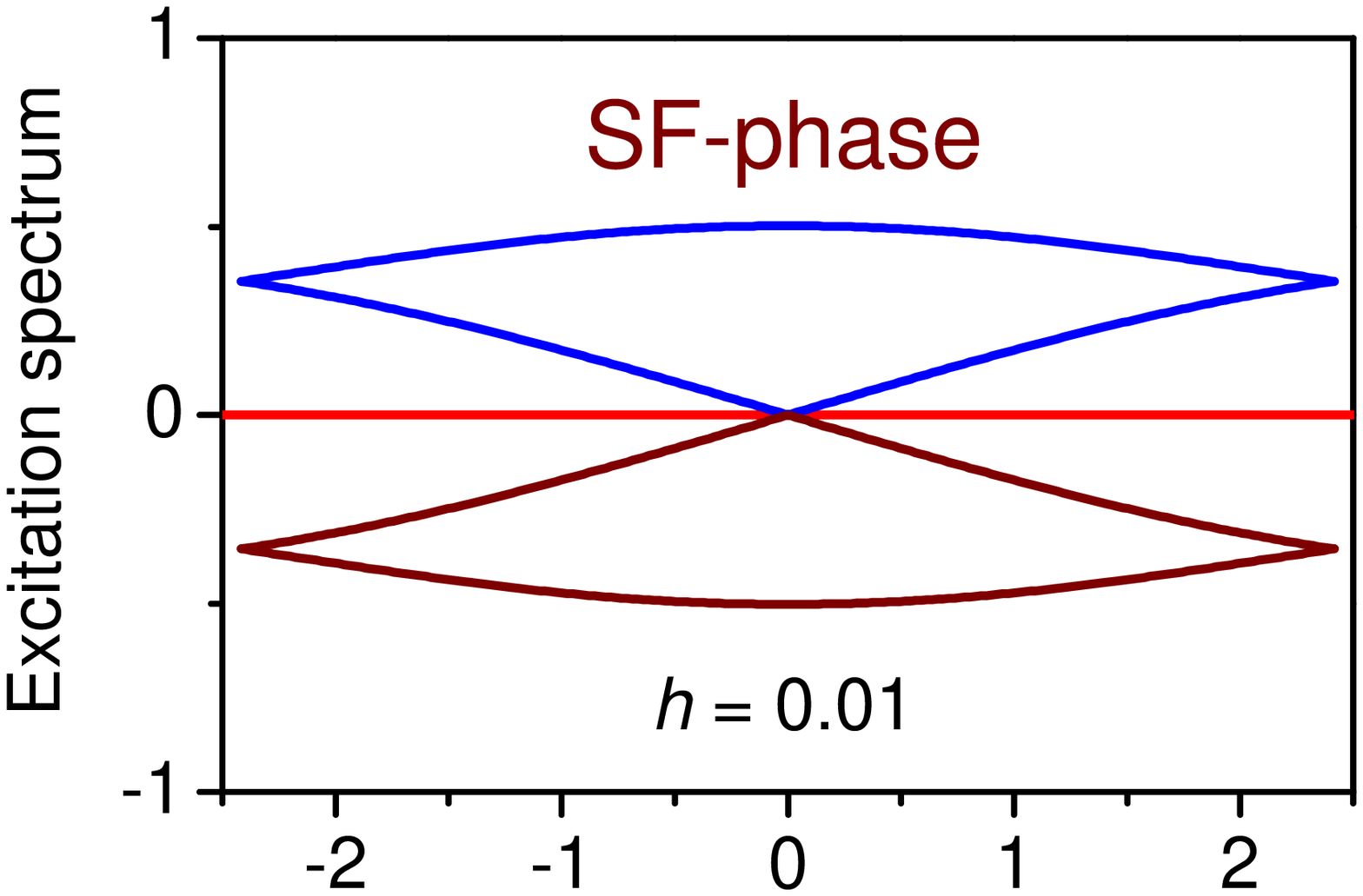}%
\hfill%
\includegraphics[height=0.15\textheight]{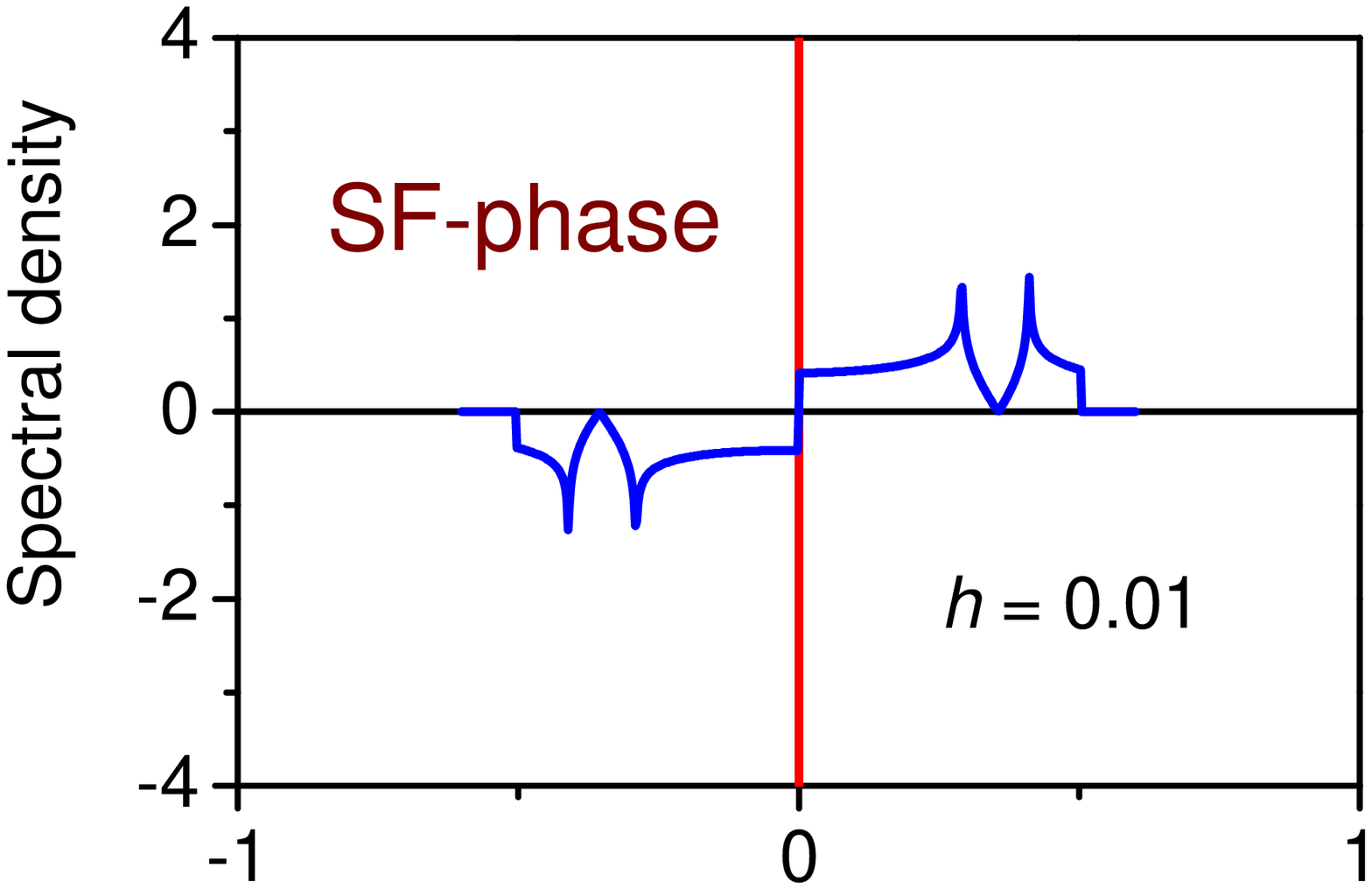}%
\hfill%
\strut%
\\[0.5mm]%
\strut%
\hfill%
\includegraphics[height=0.175\textheight]{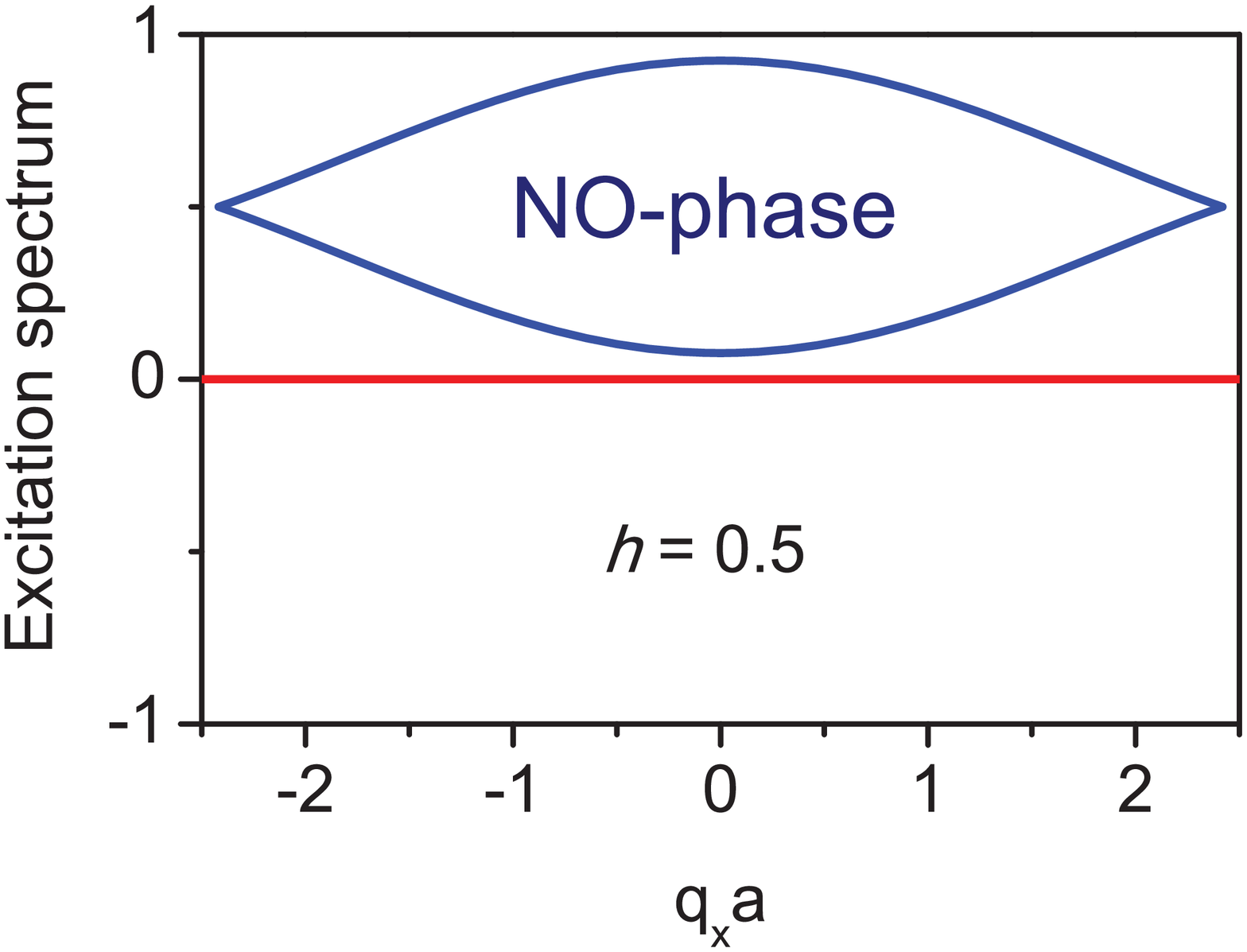}%
\hfill%
\includegraphics[height=0.175\textheight]{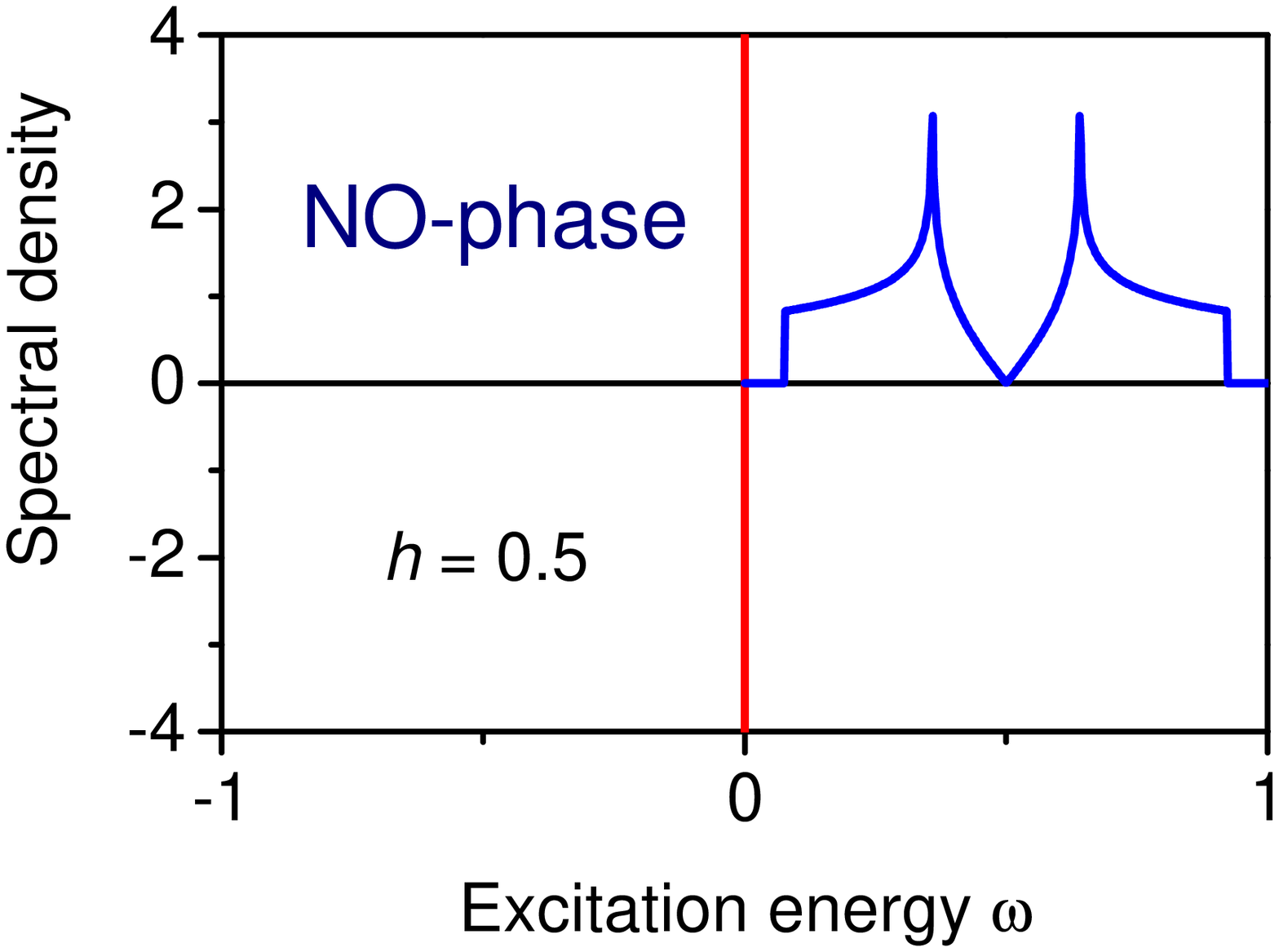}%
\hfill%
\strut%
\vskip-3mm\caption{Spectrum (left panels) and spectral density
{$\rho_{A}(\omega)$} (right panels) in the NO and SF phases at
$\delta=0$ at various energies $h$.\,\,The other parameters are $J(0)=1$ and $\Theta=0.2$}%
\label{fig06}\vspace*{-1mm}
\end{figure*}

For the $\delta$-function in formula (\ref{Pr-ista3.2}), the following
expression is used:%
\[
    \delta\left(\!\omega-\frac{\varepsilon_{i}(x)}{\hbar}\!\right)
    =
    \]\vspace*{-7mm}
    \begin{equation}
    =
    \delta(x-\tilde{x}_{i})2\hbar|\varepsilon_{i}|\left|b+\frac{d}
    {2(\hbar^{2}\omega^{2}-a-b\tilde{x}_{i})}\right|^{-1}\!\!,
    \label{Pr-ista3.6}
\end{equation}
where $\varepsilon_{i}(x)=\left.
\varepsilon_{i}(\mathbf{q})\right\vert _{|\gamma_{q}|^{2}\rightarrow
x}$, and $\tilde{x}_{i}$ is the root of the equation
$\hbar\omega=\varepsilon_{i}(x)$.\,\,We also introduced the
notations
\begin{equation}
\begin{array}{l}
     \displaystyle
    a=
    \frac{E_{A}^{2}+E_{B}^{2}}{2},
    \\[5mm]
   \displaystyle b=
    \langle\sigma_{A}^{z}\rangle\langle\sigma_{B}^{z}\rangle\cos\vartheta_{A}\cos\vartheta_{B}t^{2}
    \equiv
    m\cos\vartheta_{A}\cos\vartheta_{B},
    \\[5mm]
   \displaystyle d= m \left(E_{A}+E_{B}\cos\vartheta_{A}\cos\vartheta_{B}\right)\times
    \\[5mm]
    \displaystyle\times
    \left(E_{B}+E_{A}\cos\vartheta_{A}\cos\vartheta_{B}\right)\!.
    \label{Pr-ista3.7}
\end{array}\!\!\!\!\!\!\!\!\!\!\!\!\!\!\!\!\!\!\!\!\!\!\!\!\!\!\!\!\!\!
\end{equation}
As a result, we obtain the following expression for the spectral
density that characterizes the distribution over the
energy:\vspace*{-1mm}
\[
\rho_{A}(\hbar\omega)=\sum_{i=1}^{4}\rho_{0}(\tilde{x}_{i})\langle\sigma_{A}^{z}\rangle\,\times
\]\vspace*{-5mm}
\begin{equation}
\times
\frac{k+l\tilde{x}_{i}}{|2b(\hbar^{2}\omega^{2}-a-b\tilde{x}_{i})+d|}%
\,(-1)^{i-1}, \label{Pr-ista3.8}%
\end{equation}\vspace*{-5mm}
\begin{equation}
\begin{array}{l}
     \displaystyle
    k
    =
   \! \left[E_{A}(\cos^{2}\vartheta_{\!A}\!+1)+2\hbar\omega\cos\vartheta_{\!A}\right]\!(\hbar^{2}\omega^{2}\!-E_{B}^{2}),
    \\[1mm]
   \displaystyle l
    =
    m\cos^{2}\vartheta_{\!A}\left[(1+\cos^{2}\vartheta_{B})E_{B}-2\hbar\omega\cos\vartheta_{\!B}\right]\!.
    \label{Pr-ista3.9}
\end{array}\!\!\!\!\!\!\!\!\!\!\!\!\!\!\!\!\!\!
\end{equation}
An expression for $\rho_{B}(\hbar\omega)$ can be obtained from
formulas (\ref{Pr-ista3.8}) and (\ref{Pr-ista3.9}) by swopping the
indices: $A\rightleftarrows B$.

Formula (\ref{Pr-ista3.8}) was used in the numerical calculations of
the curves $\rho_{A}(\hbar\omega)$ describing the frequency
dependence of the spectral density for sublattice~$A$.\,\,In Figs.~4
to 6, the results obtained for various values of parameters $h$ and
$\delta$ and various temperatures are shown.\,\,As was done earlier,
the energetic quantities are reckoned in the $J(0)$-units.

A common feature of those plots is the sign change by the function
$\rho _{A}(\omega)$ when crossing the point $\omega=0$. At
$\omega<0$ (below the chemical potential level), the spectral
density is negative, and, at $\omega>0$ (above the $\mu$-level), it
is positive.\,\,The additional subbands that arise in the SF phase
possess a much smaller weight at first, and their spectral densities
can differ from the $\rho_{A}$-values for the subbands in the NO
phase even by several orders of magnitude (see Fig.~4).\,\,As the
chemical potential goes deeper into the subband, where it is located
(this process corresponds to the variation of the parameter $h$ (see
Fig.~5)), the intensity redistribution between the positive and
negative branches of the function $\rho_{A}$ takes place.\,\,The
dispersion curves $\varepsilon_{i}(\mathbf{q})$ change
insignificantly at that.

Figure~6 illustrates the reconstruction of the spectral density
function $\rho_{A}(\hbar\omega),$ when the chemical potential moves
within the band in the case where $\delta=0$.\,\,In this case, the
band is not split (the gap is absent), and there are Dirac points in
the spectrum.\,\,In the SF phase, their number doubles in comparison
with the normal one.\,\,This fact manifests itself in the appearance
of additional intermediate points in the spectral density $\rho
_{A}$, where $\rho_{A}(\hbar\omega)=0$ (in a vicinity of those
points, the frequency dependence of $\rho_{A}$ can be approximated
by a linear function).

\section{Conclusions}

Our calculations of the dispersion laws
$\varepsilon_{i}(\mathbf{q})$ and the spectral densities
$\rho_{\alpha}(\hbar\omega)$ reveal main features in the band
spectrum structure of hard-core bosons in a two-dimensional lattice
of the graphene type.\,\,The changes in their spectral
characteristics at the transition from the NO to SF phase are
described.\,\,It is shown that the form of the functions
$\rho_{\alpha}(\hbar\omega)$ ($\alpha=A,B$) is much more sensitive
to the values of energetic parameters of the system, including the
position of the chemical potential of bosons, than the dispersion
laws $\varepsilon_{i}(\mathbf{q})$ in the bosonic bands.\,\,For this
reason, the functions $\rho_{\alpha}(\hbar\omega)$ can be considered
as the basic characteristics of the band spectrum.\,\,The form of
the dependences $\varepsilon_{i}(\mathbf{q})$ alone does not provide
its exhaustive des\-cription.

It is found that, in the case of a lattice with energetically
equivalent sublattices (at $\delta=0$), the Dirac points in the
spectrum survive at the transition to the SF phase, and their number
doubles.\,\,In the general case, the chemical potential of bosons
remains outside a vicinity of the Dirac points, and it cannot be
imposed onto them.

The results obtained can form a basis for the further researches of
thermodynamic properties of the system of Bose particles in a
honeycomb lattice with the graphene-type structure.

\vspace*{-5mm}
\rezume{%
І.В.\,Стасюк, O.B.\,Величко, І.Р.\,Дулепа}{ДОСЛІДЖЕННЯ БОЗОННОГО\\
СПЕКТРА ДВОВИМІРНИХ ОПТИЧНИХ\\ ҐРАТОК ЗІ СТРУКТУРОЮ ТИПУ\\ ГРАФЕНУ.
НАДПЛИННА ФАЗА} {Досліджено енергетичний спектр системи бозе-атомів
у надплинній фазі в оптичних ґратках типу графену. Розрахунок
законів дисперсії у зонах та одночастинкових спектральних густин
проведено у наближенні хаотичних фаз у рамках формалізму жорстких
бозонів. Описано їх зміни при переході від нормальної до надплинної
фази. Під час такої перебудови збільшується вдвічі число підзон. У
випадку енергетичної еквівалентності підґраток діраківські точки у
спектрі зберігаються, а їх кількість подвоюється. При енергетичній
відмінності між підґратками точки Дірака відсутні. Показано, що
форма спектральних густин чутлива до зміни температури та
розташування хімічного потенціалу.}


\vskip2mm

\begin{thebibliography}{99}                                                                                               %

\bibitem {Stasyuk}I.V. Stasyuk, I.R. Dulepa, and O.V.~Velychko, Ukr. J.
Phys. {\bf 59}, 888 (2014).
\bibitem {Greiner02a}M. Greiner, O. Mandel, T.~Esslinger, T.W.~H{\"{a}}nsch,
and I.~Bloch, Nature \textbf{415}, 39 (2002).
\bibitem {Greiner02b}M. Greiner, O. Mandel, T.W.~H{\"{a}}nsch, and I.~Bloch,
Nature \textbf{419}, 51 (2002).
\bibitem {arr4}P. Soltau-Panahi, J.
Struck, A. Bick, W. Plenkers, G.~Mei\-neke, C. Becker, P.
Windpassinger, K. Sengstock, P.~Hau\-ke, and M. Lewenstein, Nature
Physics \textbf{7}, 434 (2011).
\bibitem {ar2}D.-S. L{\"{u}}hmann, Phys. Rev.~A \textbf{87}, 043619 (2013).
\bibitem {ar3}Q.-Q. Lu{ and} J.-M. Hou, Commun. Theor. Phys. \textbf{53}, 861 (2010).
\bibitem{ar4} P. Soltau-Panahi, D.-S. L{\"u}hmann,
    J.~Struck, P.~Wind\-pas\-sin\-ger, and K. Sengstock,
    Nature Physics \textbf{8}, 71 (2012).
\bibitem {ar5}E. Albus, X. Fernandez-Gonzalvo, J.~Mur-Petit,
J.J.~Gar\-cia-Ri\-poli,{ and} J.K.~Pachos, Ann. Phys. \textbf{328},
64 (2013).
\bibitem {Fisher89}M.P.A. Fisher, P.B. Weichman, G.~Grinstein,{ and}
D.S.~Fi\-sher, Phys. Rev.~B \textbf{40}, 546 (1989).
\bibitem {Jaksch98}D. Jaksch, C. Bruder, J.I.~Cirac, C.W.~Gardiner,{ and}
P.~Zol\-ler, Phys. Rev. Lett. \textbf{81}, 3108 (1998).
\bibitem {a6}Z. Chen{ and} B. Wu, Phys. Rev. Lett. \textbf{107}, 065301 (2011).
\bibitem {a7}S. Koghee, L.-K. Lim, M.O.~Goerbig,{ and} C.~Morais-Smith, Phys.
Rev.~A \textbf{85}, 023637 (2012).
\bibitem {Whitlock63}R.T. Whitlock{ and} P.R.~Zilsel, Phys. Rev. \textbf{131},
2409 (1963).
\bibitem {a1}I.V.~Stasyuk{ and} O.~Vorobyov, Condens. Matter Phys.
\textbf{16}, 23005 (2013).
\bibitem {a2}A.H.~Castro Neto, F.~Guinea, N.M.R.~Peres, K.S.~No\-vo\-se\-lov, and
A.K.~Geim, Rev. Mod. Phys. \textbf{81}, 109 (2009).
\bibitem {a3}N.N. Bogolyubov, Izv. Akad. Nauk SSSR Ser. Fiz. \textbf{11}, 77 (1947).
\bibitem {a4}H.B.~Rosenstock, J.~Chem. Phys. \textbf{21}, 2064 (1953).
\bibitem {a5}J.P.~Hobson{ and} W.A.~Nierenberg, Phys. Rev. \textbf{89}, 662 (1953).\vspace*{-2mm}
\begin{flushright}
{\footnotesize Received 10.06.14.\\ Translated from Ukrainian by
O.I.~Voitenko}
\end{flushright}
\end{thebibliography}
\end{document}